# Materials discovery and design principles for stable, high activity perovskite cathodes for solid oxide fuel cells

Ryan Jacobs,[1] Tam Mayeshiba,[1] John Booske[2], and Dane Morgan[1,*]

[1]Department of Materials Science and Engineering, University of Wisconsin- Madison, Madison, WI.

[2]Department of Electrical and Computer Engineering, University of Wisconsin- Madison, Madison, WI.

*Corresponding author e-mail: ddmorgan@wisc.edu

†Electronic supporting information (SI) available

## Abstract

Critical to the development of improved solid oxide fuel cell (SOFC) technology are novel compounds with high oxygen reduction reaction (ORR) catalytic activity and robust stability under cathode operating conditions. We have screened approximately 2145 distinct perovskite compositions for potential use as high activity, stable SOFC cathodes, and have verified that our screening methodology qualitatively reproduces the experimental activity, stability, and conduction properties of well-studied cathode materials. We used the calculated oxygen *p*-band center as a first principles-based descriptor of the surface exchange coefficient ($k^*$), which in turn correlates with cathode ORR activity. We used convex hull analysis under operating conditions in the presence of oxygen, hydrogen, and water vapor to determine thermodynamic stability. This search has yielded 52 potential cathode materials with good predicted stability in typical SOFC operating conditions and predicted $k^*$ on par with leading ORR perovskite catalysts. We also used our established trends in predicted $k^*$ and stability to suggest methods of improving the performance of known promising compounds. The materials design strategies and new materials discovered in our computational search will help enable the development of high activity, stable compounds for use in future solid oxide fuel cells and related applications.

## Table of Contents Figure

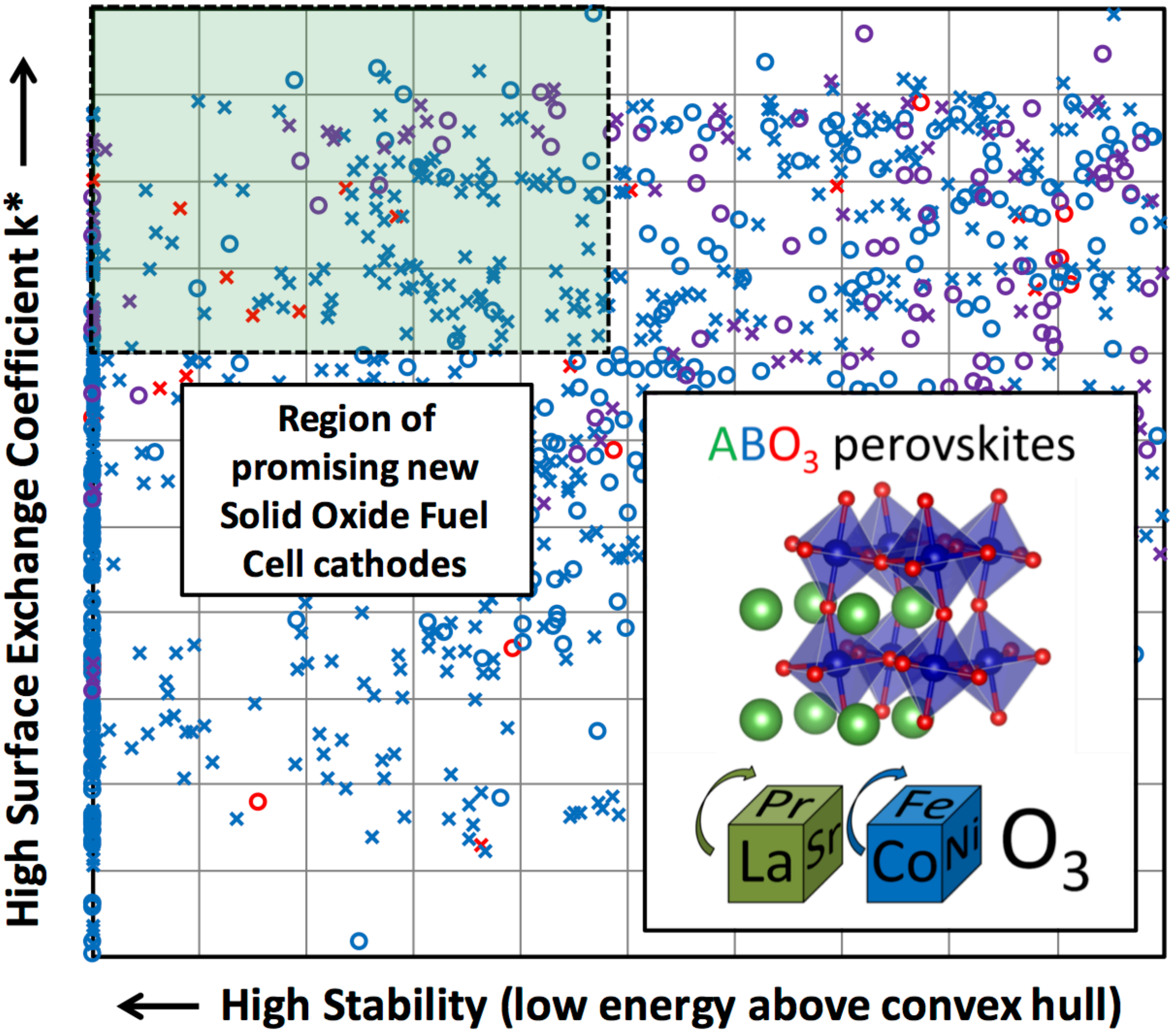

# 1. Introduction

The increasing demand to reduce the dependency on fossil fuels has necessitated advancements in device-related materials for alternative energy technologies. Solid oxide fuel cells (SOFCs) may play an important role in the future of energy technology, as they are able to produce energy by direct chemical-to-electrical conversion of oxygen and hydrogen or hydrocarbon fuels with high efficiency and relatively little emission of greenhouse gases.[1, 2] When operated in reverse, the fuel cell functions as an electrolyzer, effectively storing the energy obtained by splitting water into hydrogen and oxygen for future power generation.[3]

Typically, SOFCs must be operated at high temperatures of around 800-1000 °C, primarily in order to overcome the slow kinetics of the oxygen reduction reaction (ORR) ($O_2 + 4e^- \rightarrow 2O^{2-}$) at the cathode, which results in a high cathodic overpotential at lower temperature.[4, 5] Even at higher temperatures the slow kinetics of the ORR is a major contributor to the overall resistance of the SOFC, resulting in decreased device efficiency.[4] High temperature operation of the SOFC causes accelerated materials degradation and high operational costs. A key materials property that correlates with the ORR is the surface exchange coefficient $k^*$, as higher $k^*$ values correspond to more rapid splitting of the $O_2$ and incorporation into the cathode, which in turns correlates with more efficient overall ORR. A cathode with a higher value of the surface exchange coefficient $k^*$ and correspondingly improved ORR activity would allow for lower temperature operation of the SOFC, which in turn improves fuel cell lifetime by slowing materials degradation. Such cathode improvements would increase the economic incentive for large scale commercialization of SOFC technology.[5]

Perovskite oxides have presented themselves as the most promising alternative materials to precious metal alloys for SOFCs,[6-8] with materials such as $La_{1-x}Sr_xMnO_3$ (LSM) and $La_{1-x}Sr_xCo_{1-y}Fe_yO_3$ (LSCF) appearing widely in commercial SOFCs.[8-10] Similar perovskites also appear to be very promising for aqueous electrocatalysis, with materials such as $Ba_{1-x}Sr_xCo_{1-y}Fe_yO_3$ (BSCF) and $Pr_{1-x}Ba_xCoO_3$ (PBCO) displaying record high ORR[11] and oxygen evolution reaction (OER)[12] activities, respectively. While BSCF appears to have a high activity for both high temperature ORR and room temperature aqueous OER, it unfortunately suffers from stability problems.[13-15] This work seeks to discover high activity perovskite cathode materials that are also stable under SOFC operating conditions.

We have used high-throughput Density Functional Theory (DFT) methods enabled by the MAterials Simulation Toolkit (MAST),[16] combined with multicomponent phase stability analysis using the Python Materials Genomics (Pymatgen)[17] toolkit, to screen a wide composition space of perovskite materials for candidates that have high $k^*$ and are stable under typical SOFC operating conditions. As a reference for the remainder of this work, we will use the term "ORR operating conditions" to signify high-temperature SOFC cathodic working conditions, where $T$ = 800 °C (1073 K), $p(O_2)$ = 0.21 atm and $H_2O$ gas is present with a typical relative humidity (*RH*) of *RH* = 30%.[18-20] The main focus of this study was on perovskite materials, which have the chemical formula $A_{1-x}A'_xB_{1-y}B'_yO_3$. A subset of the most promising perovskite materials identified in this work were further modeled as Ruddlesden-Popper and hexagonal phases (see **Figure 1**) to assess the stability of the perovskite phase in the presence of these competing phases. In total, 2145 distinct perovskite compositions were simulated. **Figure 1** depicts the different structures modeled in this work. Additional details regarding the modeling of these doped perovskite structures can be found in **Section 5** of the **Supporting Information (SI)**.

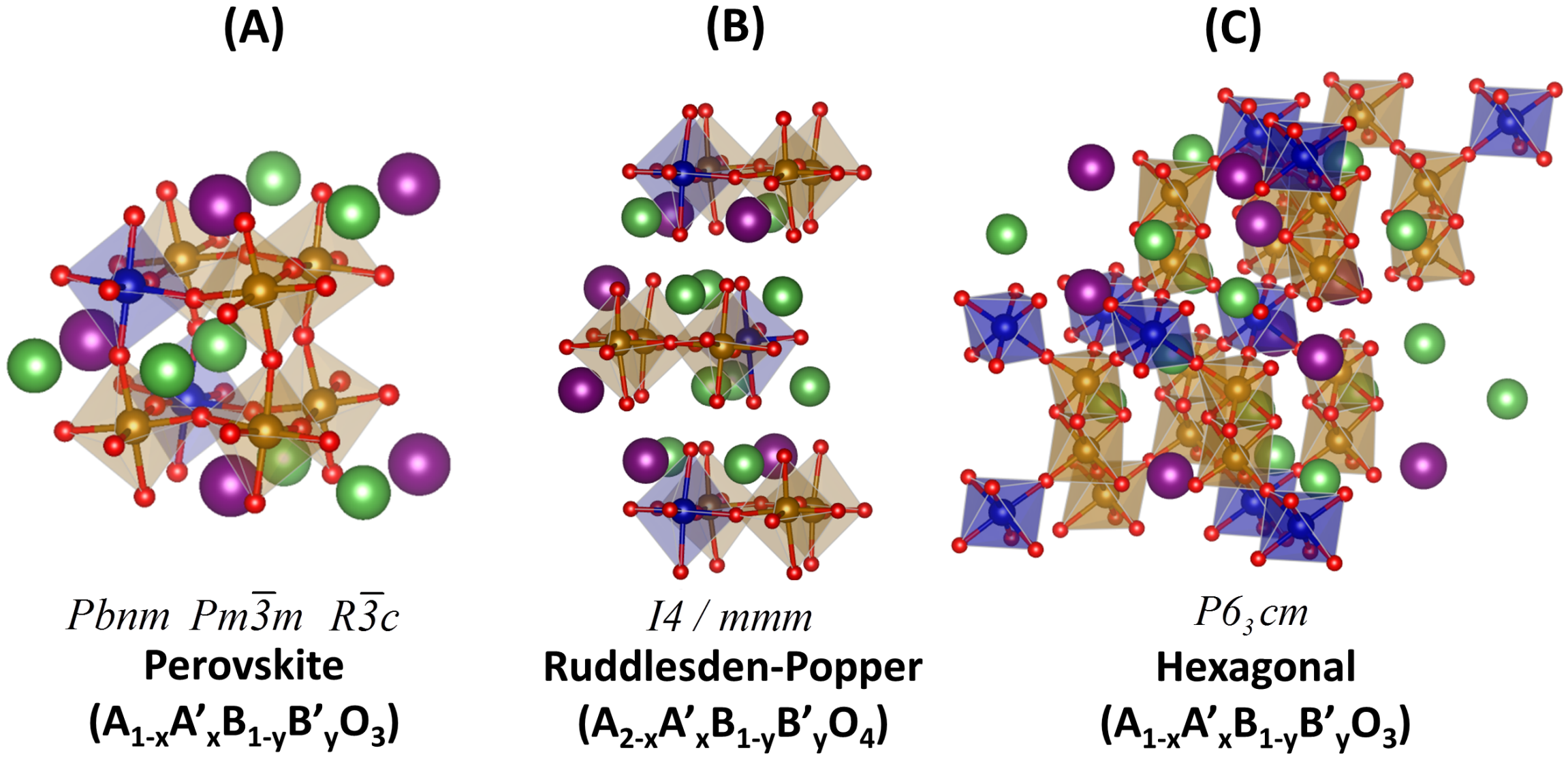

**Figure 1.** Structures of materials simulated in this work. (A) Perovskite structure, which has either orthorhombic $Pbnm$, rhombohedral $R\bar{3}c$ or cubic $Pm\bar{3}m$ space groups. (B) Ruddlesden-Popper n=1 structure with tetragonal space group $I4/mmm$. (C) Hexagonal structure with $P6_3cm$ space group. For all structures, the specific material shown here is $La_{0.625}Sr_{0.375}Co_{0.25}Fe_{0.75}O_3$ (LSCF). The green, purple, blue, gold, and red spheres signify La, Sr, Co, Fe and O atoms, respectively.

It has been shown recently that the calculated value of the O (2)*p*-band center (our approach to calculating this quantity from the density of states is given in **Section 4**) is a good descriptor for experimental high-temperature surface exchange coefficients for a wide range of perovskite materials.[12, 21, 22] This descriptor has also been shown to correlate with numerous oxide properties, including: point defect energies (O vacancies and interstitials),[21, 23, 24] activation energies for diffusion and surface exchange,[22, 25] the values of log($k^*$) and log(ASR) (ASR= area specific resistance),[21] work function,[26] and low temperature OER,[12] suggesting it plays a critical role in many oxide properties. Here, we use this descriptor to screen a large number of potential compounds to (1) discover new potential perovskite ORR cathodes which have a high value of the O $p$-band center and thus maximize the predicted $k^*$, and (2) understand which alloying elements act to maximize the predicted $k^*$ and subsequently narrow our materials search to these elements. All materials were subject to a full analysis of chemical stability under typical ORR operating conditions to assess their overall applicability as new cathodes.

**Figure 2** summarizes the sequential steps of our screening process and elimination criteria. First, we generated and simulated a wide composition space of perovskite materials. Overall, 72 ternary, 1359 quaternary, and 714 quinary perovskites were investigated, for a total of 2145 distinct perovskite oxide compositions. For additional details on the specific composition ranges and elements alloyed on the A- and B-sites, see **Section 4** of the **SI**. In addition, a full list of every material composition modeled in this work is tabulated in a spreadsheet as part of the **SI**. We analyzed the predicted surface exchange coefficient $k^*$ using the O $p$-band center as a descriptor for $k^*$ (see **Section 2.1**), and eliminated materials that had a predicted value of $k^*$ less than the experimentally measured $k^*$ value of $La_{0.625}Sr_{0.375}Co_{0.25}Fe_{0.75}O_3$ (LSCF), which is already a commercial cathode material for SOFCs. Next, we analyzed the chemical stability under ORR operating conditions using the phase stability analysis tools contained in Pymatgen (see **Section 2.3**), and any materials with a predicted instability greater than 47 meV/atom above the convex hull were eliminated from consideration. This value of 47 meV/atom was chosen because it is the calculated stability of the commercial material LSCF, which has demonstrated good stability over a long operating time and thus represents a logical stability limit. Furthermore, an uncertainty of 40 meV/atom resulting from typical DFT errors is a conservative estimate to include in the screening.[27-29] We note that our cutoff of 47 meV/atom is reasonably close to this 40 meV/atom DFT error bar. Our final elimination criterion was to remove any compounds that possessed a nonzero bandgap or charge transfer gap, thus eliminating materials which would likely be poor conductors. As **Figure 2** illustrates, the application of each successive elimination criterion reduces the total pool of potentially promising material candidates. After our screening process was completed, we were left with a small pool of 52 promising perovskite materials as new SOFC cathodes. The computational methods used to calculate the values used in this screening are described in **Section 4**. For select cases, we also explored the stability of the cathode with competing Ruddlesden-Popper and hexagonal phases (details can be found in **Section 11** of the **SI**) and commonly used electrolytes (details can be found in **Section 12** of the **SI**). We note here that in addition to the screening criteria detailed in **Figure 2**, thermal expansion mismatch between the cathode and electrolyte is also an important SOFC design parameter, and is calculable with DFT methods.[30-33] However, a detailed analysis of the thermal properties of these materials is beyond the scope of the current study, and may be evaluated for a targeted set of promising materials in a future study.

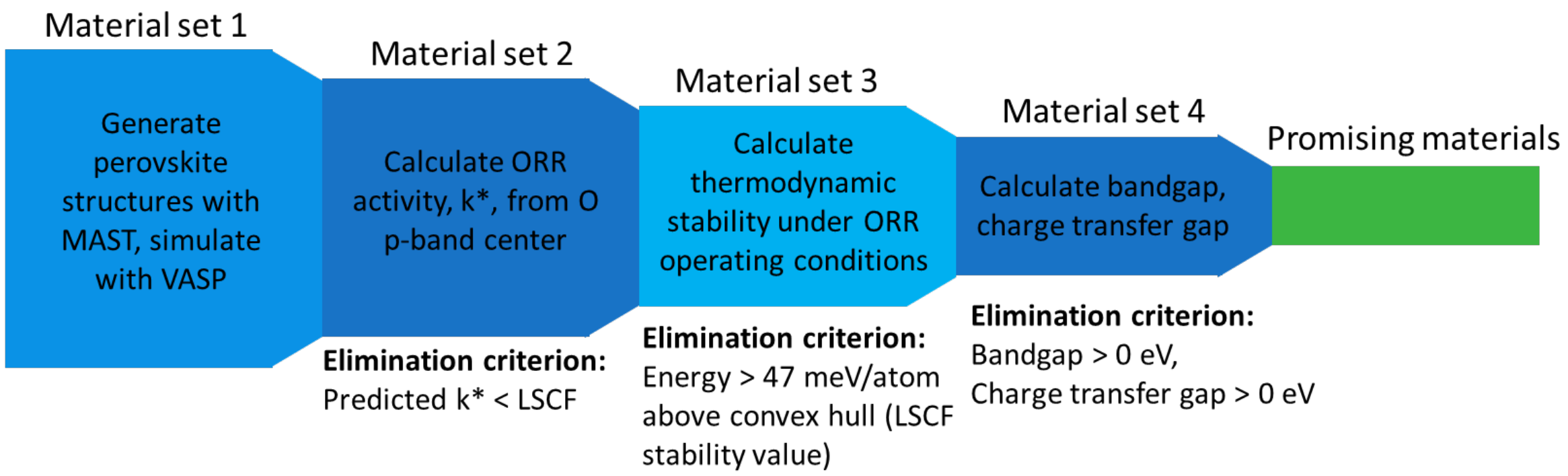

**Figure 2.** Summary of screening and elimination criteria used in this study. In stages (1) through (4), the listed elimination criteria are invoked to reduce the number of potential candidate materials. By stage (5), only a small number of potential candidate materials remain. These materials have predicted surface exchange higher than the commercial material LSCF, stability better than 47 meV/atom above the convex hull, and possess electronic structures with zero bandgap and zero charge transfer gap.

# 2. Results and Discussion

## 2.1. Predicted surface exchange from the O *p*-band center descriptor

To obtain a quantitative relationship between the calculated bulk O *p*-band center and the experimental surface exchange coefficient $k^*$, we fit our calculated bulk O *p*-band center values for a series of perovskite materials whose experimental surface exchange coefficients have been measured or can be readily estimated at T ≈ 1000 K and $p(O_2)$ ≈ 0.2 atm (**Figure 3**). These experimental data were obtained from Ref. [21] and references therein, and the references provided in the caption of **Figure 3**. The materials shown with green diamonds are promising new cathode materials which passed all screening elimination criteria (see **Table 1** of **Section 2.3** and the data spreadsheet as part of the **SI** for the list of these materials). These data were plotted using the line of best fit shown in **Figure 3**. In this work we included in **Figure 3** thirteen points in addition to the original nine from Ref. [21], which all together strongly support the linear trend originally identified in Ref. [21], although with a slightly different linear relationship.

In addition to the data used in the fit, we further expanded our set of $k^*$ data by including some systems where no $k^*$ measurement was available but where we were able to find measurements of area-specific resistance (ASR) ($La_{0.2}Sr_{0.4}Ba_{0.4}Fe_{0.875}Mn_{0.125}O_3$ (LSBFM)[34], $BaFeO_{2.5}$ (BF)[35] and $BaCo_{0.625}Fe_{0.25}Nb_{0.125}O_3$ (BCFN)[36]). We then estimated $k^*$ for these systems from the ASR and the linear correlation $\log(k^*) = -1.3031\times\log(ASR) - 5.9496$, which we obtained by fitting to the $k^*$ and ASR data from Ref. [21]. A plot of the $k^*$ versus ASR data is provided in **Figure S3** in **Section 8** of the **SI**. This data was not used in fitting $k^*$ vs. O $p$-band center due to its uncertainty but is shown in **Figure 3 (**purple diamonds) and further corroborated the established trend of $k^*$ versus O $p$-band center.

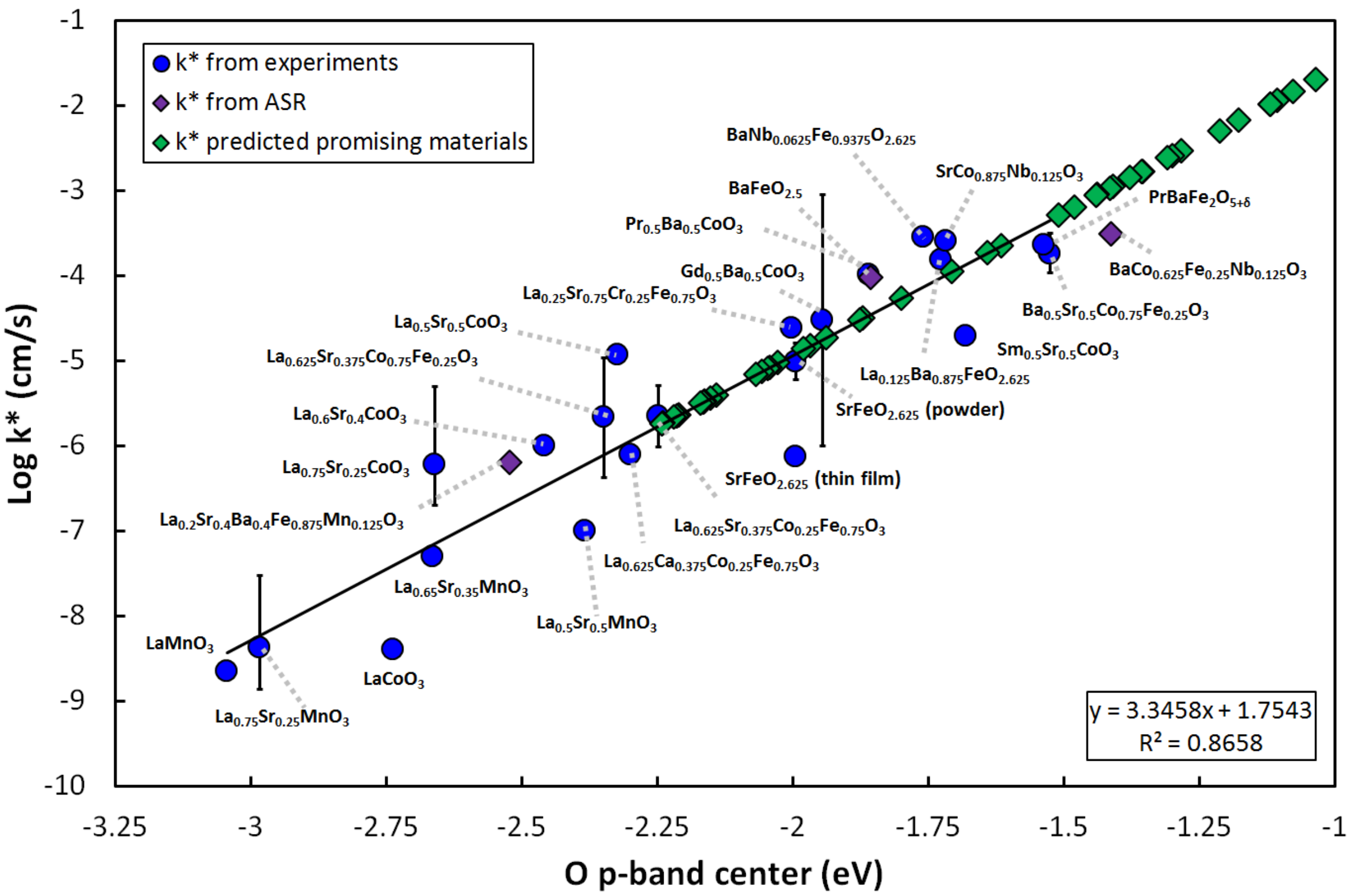

**Figure 3.** Plot of surface exchange coefficient log $k^*$ (in cm/s) as a function of calculated O $p$-band center (in eV). The O $p$-band center values are given relative to the Fermi energy for each material. The blue symbols are experimental surface exchange coefficient data. The line of best fit was made from the materials comprising the blue circles, which were obtained from Ref. [21], and the following materials and references: $La_{0.6}Sr_{0.4}CoO_3$: Ref. [37], $La_{0.5}Sr_{0.5}CoO_3$: Refs. [38, 39], $La_{0.25}Sr_{0.75}Cr_{0.25}Fe_{0.75}O_3$ (LSCrF): Ref. [40], $SrFeO_{2.625}$ (SF): Ref. [40] (powder), Ref. [41] (thin film), $La_{0.0625}Ba_{0.9375}FeO_{2.625}$ (LBF): Ref. [42], $BaNb_{0.0625}Fe_{0.9375}O_{2.625}$ (BFN): Ref. [43], $PrBaFe_2O_{5+d}$ (PBFO): Ref. [44], and $SrCo_{0.875}Nb_{0.125}O_3$ (SCN): Ref. [45]. The $k^*$ data for the purple symbols was obtained by using ASR data reported for $La_{0.2}Sr_{0.4}Ba_{0.4}Fe_{0.875}Mn_{0.125}O_3$ (LSBFMO): Ref. [34], $BaFeO_{2.5}$ (BF): Ref. [35], and $BaCo_{0.625}Fe_{0.25}Nb_{0.125}O_3$ (BCFN): Ref. [36]. The green diamonds were plotted using predicted log $k^*$ values based on the linear fit of the experimental data. Some of the references listed above reported values of $k_{chem}$, which is related to $k^*$ through the relation $k^* = k_{chem} x_{V_O}$, where $x_{V_O}$ is the concentration of oxygen vacancies in the material determined following Ref. [46]. The microstructure of all materials present in this plot are dense polycrystalline pellets, except BSCF (polycrystalline PLD film), $LaCoO_3$ (single crystal), SF (film) (polycrystalline thin film), LBF and BFN (both spray deposited polycrystalline powders). The error bars were made by taking the maximum and minimum reported $k^*$ values (the fitted data points are the average $k^*$ values) from multiple references: $La_{0.75}Sr_{0.25}MnO_3$: Refs. [47-49], $La_{0.75}Co_{0.25}O_3$: Refs. [47, 50], $La_{0.625}Sr_{0.375}Co_{0.75}Fe_{0.25}O_3$: Refs. [51, 52], $La_{0.625}Sr_{0.375}Co_{0.25}Fe_{0.75}O_3$: Refs. [51, 53], $SrFeO_{2.625}$: Refs. [40, 41], $Gd_{0.5}Ba_{0.5}CoO_3$: Refs. [54,

55], $Ba_{0.5}Sr_{0.5}Co_{0.75}Fe_{0.25}O_3$: Refs. [56-59]. All materials included here were calculated to be either metallic or *n*-type semiconductors.

The materials LBF[42], BFN[43], and PBFO[44] all have experimental $k^*$ values which are similar to or exceed BSCF. From the current screening study, the materials LBF, BFN and PBFO all have high predicted $k^*$ values. However, all of LBF, BFN, PBFO and BSCF failed the screening criteria related to stability or band/charge transfer gap. These findings suggest that while LBF, BFN and PBFO are potentially promising materials for SOFC cathodes, they may either suffer from long-term stability problems if they were to be in operation for thousands of hours, or, if they are insulating, they may be unable to transport electrons sufficiently fast to function as a good electrode material. Encouragingly, some of the promising materials we identify in **Section 2.3** and list in **Table 1** are closely related in composition to materials like LBF, BFN, PBFO and BSCF, but with additional dopant elements that we predict will render these new promising materials more stable and/or more conductive than their undoped counterparts.[42, 43, 60-63]

All of the materials contained in **Figure 3** are intentionally chosen to be metallic, *n*-type conductors at ORR operating conditions. In addition to the materials in **Figure 3**, we have made an additional plot of $k^*$ versus O *p*-band center that also includes *p*-type conductors (see **Figure S1** of **Section 8** of the **SI**). While the *p*-type conductors qualitatively follow a similar trend as the *n*-type materials, they show significantly poorer agreement with the linear correlation (compared to the fit line the root mean squared errors (RMSE) in log $k^*$ are 0.58 and 1.63 for *n*-type and *p*-type systems, respectively. Even if one refits the line with all the data the RMSE values are 0.84 and 0.91 for the *n*-type and *p*-type systems, respectively) We propose that the worse correlation of the O *p*-band center for *p*-type systems is to be expected as the surface exchange in these materials will be significantly impacted by the rate of electronic charge transfer, and not just ionic transfer of oxygen, as shown in the work of Tuller, et al.[41] While the O *p*-band can serve as a powerful descriptor for oxygen ion energetics, it is not clear how it might correlate, if at all, with properties relevant for electron transfer limited processes. Exploring such correlation is an interesting topic for future study, but at this point we note that one should be very cautious using the correlation of $k^*$ with O *p*-band for *p*-type conductors. In addition to the O *p*-band center acting as a descriptor for $k^*$, we have shown that the relationship between experimental values of ASR and O *p*-band center (see **Figure S2** in **Section 8** of the **SI**) is also linear, and hence the O *p*-band center is a good descriptor for the ASR as well.

As a validation check for our modeling, we have compared the experimental and predicted values of $k^*$ and ASR for a set of well-studied SOFC cathode materials: LSM, LSC, LSCF and BSCF. A summary of these data can be found in **Table S2** in **Section 9** of the **SI**. In addition to the $k^*$ and ASR data, we compared our calculated stability values for these materials (see **Section 2.3** for the stability analysis) with the experimentally known qualitative behavior of the performance of these cathode materials over time in an SOFC. Finally, we have also included our calculated bandgaps compared with the experimentally known qualitative electrical conduction characteristics of these materials under SOFC operating conditions. Our methods reproduce the known characteristics of these well-studied cathode materials, which gives us confidence that our methods will provide accurate information regarding the predicted performance of new promising materials.

### 2.2. Effect of composition on O *p*-band center and stability

To understand how altering the composition of a perovskite material quantitatively affects the value of the O *p*-band center, and thus, the predicted value of the surface exchange coefficient $k^*$ (this correlation is detailed in **Section 2.1**), we show in **Figure 4** the variations in O *p*-band as a function of A-site element (for fixed B-site element) and as a function of B-site element (for fixed A-site element) for our set of ternary, undoped perovskites. Generally, one wants to maximize the value of the O *p*-band center, as a higher (less negative) O *p*-band center correlates to a higher value of $k^*$. From inspecting **Figure 4A**, it is evident that materials containing a rare earth element on the A-site generally have lower values of the O *p*-band center while materials with an alkaline earth element on the A-site have higher O *p*-band centers. This trend occurs because replacing a rare earth element with an alkaline earth results in more oxidation of the system, which in turn causes the Fermi level to move down in energy. Assuming the O *p*-band is rigid, the O *p*-band center will become higher (closer to the Fermi level) as a result. In addition, for fixed A-site, the value of the O *p*-band center tends to increase as the B-site element proceeds from left to right across the periodic table. This trend occurs because late transition metals at the right end of the periodic table are more electronegative; hence, they have a deeper *d*-band that is more hybridized with the O *p*-band compared to early transition metals. The deeper *d*-band and increased hybridization cause the Fermi level to move down in energy. Again assuming the O *p*-band is rigid, the O *p*-band center will become higher (closer to the Fermi level). Finally, materials which are band insulators such as $LaScO_3$, $PrGaO_3$, etc. also have high O *p*-band centers. The high O *p*-band center of these band insulators is the result of an empty *d*-band, which causes the Fermi level to be comprised of only a very narrow O *p*-band. As the O *p*-band is narrow, its band center is thus very close to the Fermi level, resulting in a high O *p*-band center. However, while highly insulating materials like $LaScO_3$ have a high O *p*-band center, they are not expected to be good ORR catalysts due to their insufficient electrical conductivity. Similar trends are also apparent in **Figure 4B**. It is clear that if one wants to maximize the value of the O *p*-band center, and therefore the value of $k^*$, it is generally advantageous to have alkaline earth elements such as Ca, Sr and Ba on the A-site and late transition metal elements such as Mn, Fe, Co and Ni on the B-site. These observations are consistent with the general trends seen in the literature for high-$k^*$ materials.

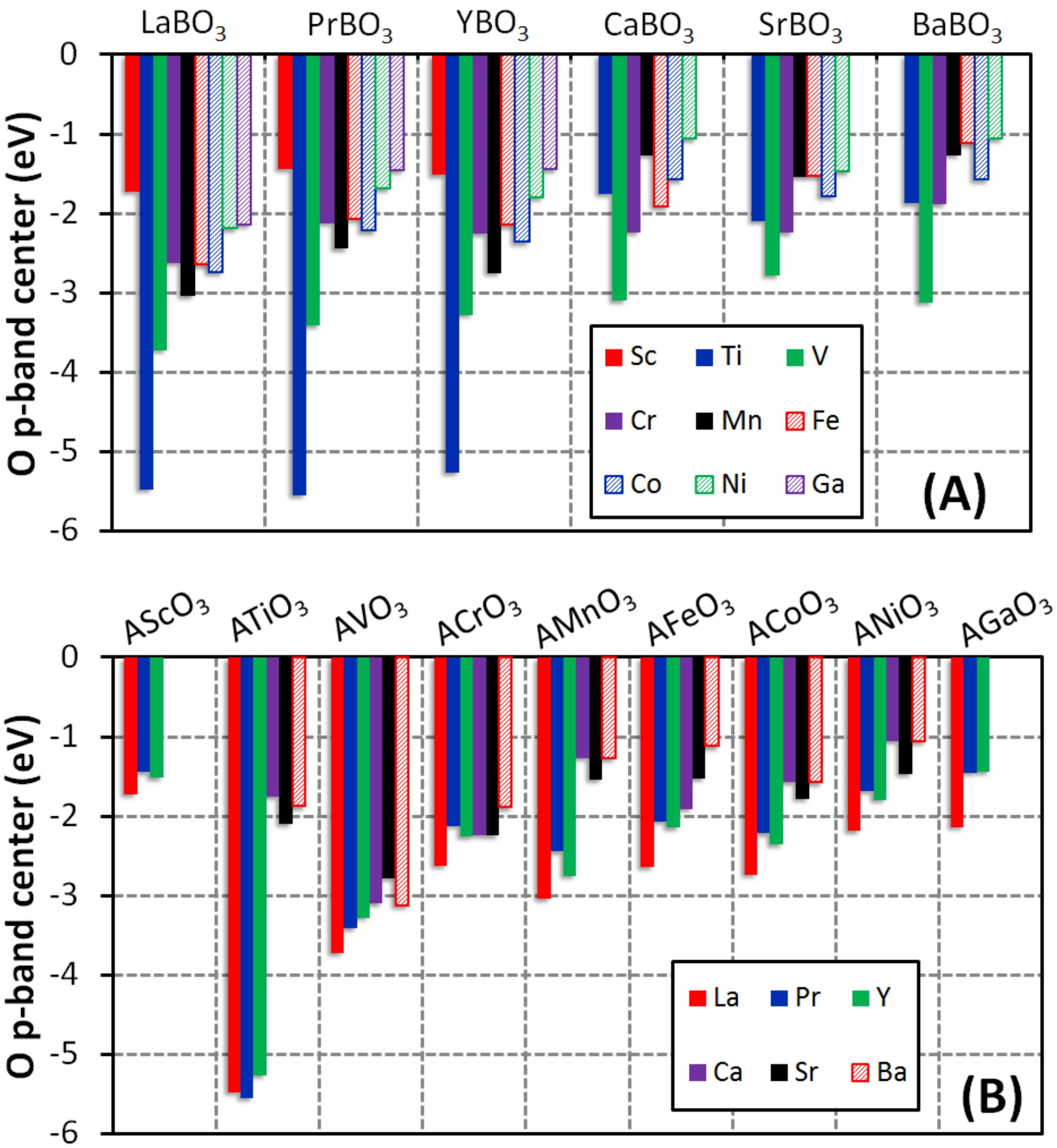

**Figure 4.** Trend of calculated O *p*-band center as a function of composition for materials with (A) different B-site elements with constant A-site element and (B) different A-site elements with constant B-site element. A high value of the O *p*-band center correlates with a high *k** value, which is indicative of high ORR activity.

In an analogous manner to our analysis of trends in O *p*-band center, we have analyzed the thermodynamic stability under typical ORR operating conditions as a function of composition for the same set of ternary perovskite compounds. **Figure 5** is a summary of the trends of stability as a function of changing B-site element with constant A-site element (**Figure 5A**), and stability as a function of changing A-site element with constant B-site element (**Figure 5B**). From **Figure 5A**, materials containing a rare earth element on the A-site tend to exhibit greater stability compared to materials containing an alkaline earth element on the A-site. In addition, for fixed A-site, the energy above the convex hull increases (stability decreases) as the B-site element proceeds from left to right across the periodic table. **Figure 5B** reinforces these stability trends, and it is particularly clear that late transition metal elements such as Fe, Co and Ni result in destabilization of perovskites under ORR conditions. These basic stability trends make sense if one considers the ease of oxidizing/reducing these elements by inspecting the trend of formation energies of binary oxides containing alkaline earths, rare earths, and transition metals. For example, based on experimental formation enthalpies tabulated on the Materials Project website,[64] the formation enthalpies of the alkaline earth oxides CaO, SrO and BaO are -6.582, -6.136 and -5.681 eV/(metal cation), respectively, whereas the formation enthalpies of rare earth oxides $La_2O_3$, $Y_2O_3$, and $Pr_2O_3$ are -9.302, -9.872 and -9.378 eV/(metal cation), respectively. Qualitatively, inclusion of rare earths can form a more stable oxidized material, whereas materials containing alkaline earths will be more easily reduced at high temperature. Analogously, the formation enthalpies of transition metal oxides $Sc_2O_3$, $TiO_2$, $V_2O_5$, $Mn_2O_3$, $Fe_2O_3$, $Co_3O_4$ and NiO are -9.889, -9.784, -8.034, -4.965, -4.278, -3.144 and -2.484 eV/(metal cation), respectively. That is, transition metals on the right side of the periodic table do not form strongly oxidized compounds as easily, which directly corresponds to the perovskites containing Co and Ni being less stable than those containing early transition metals like Sc and Ti. Furthermore, in **Figure 5C**, we have plotted the fraction of materials whose predicted $k^*$ exceeds LSCF for different ranges of energy above the convex hull, using all 2145 materials screened in this work. **Figure 5C** clearly demonstrates the above intuition that higher activity compounds tend to result in higher energy above the convex hull (i.e., lower stability). On average, if one wants to maximize the stability of perovskites, it is best to include rare earth elements on the A-site and early transition metal elements on the B-site. The highly insulating materials are also very stable, but as discussed above relating to **Figure 4**, these are not expected to be good ORR catalysts.

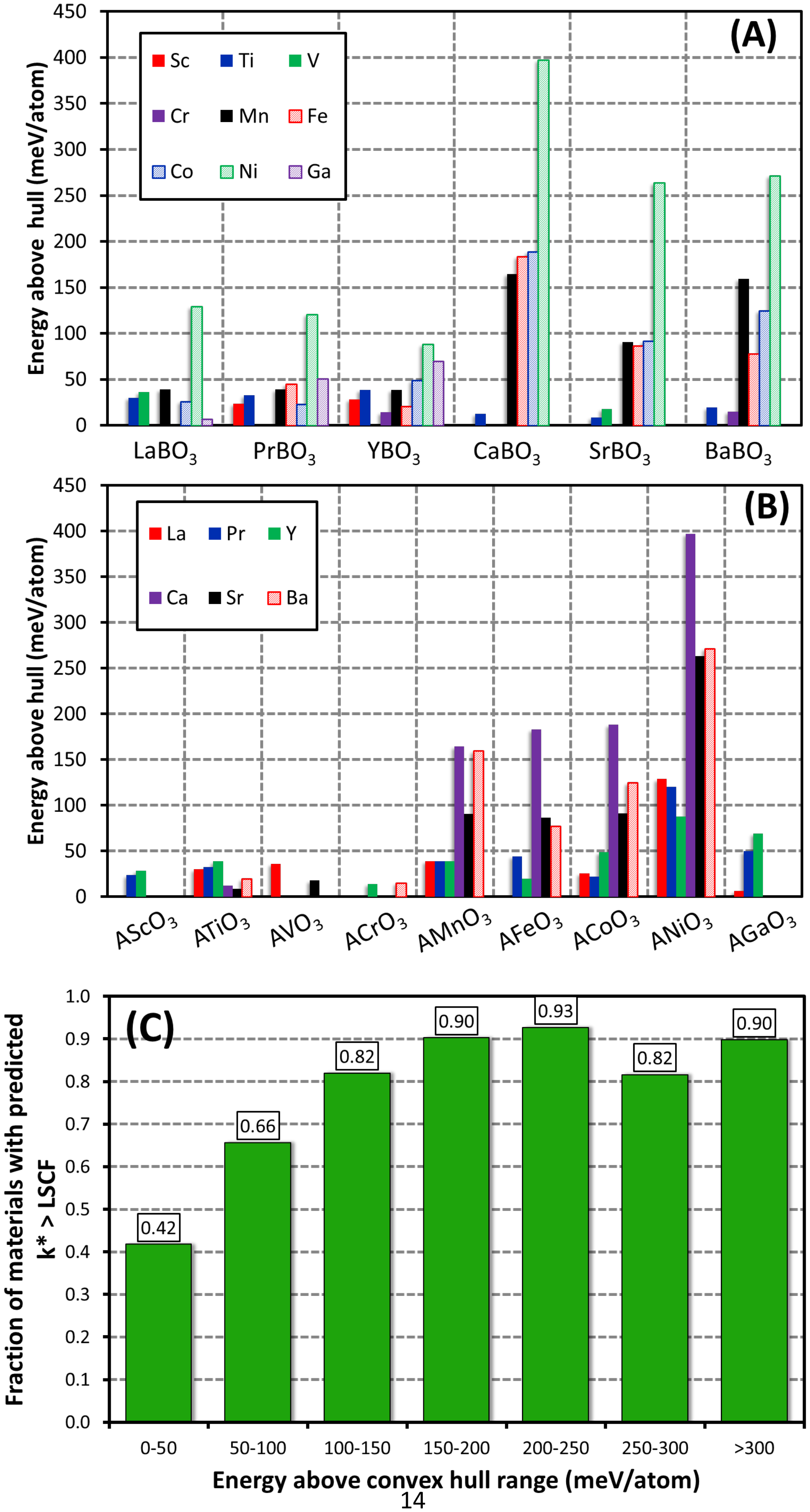
(A)
Sc
Ti
V
Cr
Mn
Fe
Co
Ni
Ga
Energy above hull (meV/atom)
450
400
350
300
250
200
150
100
50
0
$LaBO_3$
$PrBO_3$
$YBO_3$
$CaBO_3$
$SrBO_3$
$BaBO_3$
(B)
La
Pr
Y
Ca
Sr
Ba
Energy above hull (meV/atom)
$AScO_3$
$ATiO_3$
$AVO_3$
$ACrO_3$
$AMnO_3$
$AFeO_3$
$ACoO_3$
$ANiO_3$
$AGaO_3$
(C)
Fraction of materials with predicted k* > LSCF
1.0
0.9
0.8
0.7
0.6
0.5
0.4
0.3
0.2
0.1
0.0
0.42
0.66
0.82
0.90
0.93
0.82
0.90
0-50
50-100
100-150
150-200
200-250
250-300
>300
Energy above convex hull range (meV/atom)

**Figure 5.** Trend of calculated stability (as energy above the convex hull) under typical ORR operating conditions as a function of composition for materials with (A) different B-site elements with constant A-site element and (B) different A-site elements with constant B-site element. A low value of the energy above the convex hull indicates higher thermodynamic stability, which is indicative of longer SOFC cathode lifetime since the material is less prone to phase decomposition. Plot (C) catalogues the fraction of all 2145 screened materials whose predicted *k** values exceed LSCF as a function of different stability ranges, showing that higher energies above the convex hull (i.e., less stable materials) tend to result in higher activity.

The compositional trends in O *p*-band center and stability depicted in **Figure 4** and **Figure 5** are useful as they allow us to focus our subsequent materials screening on a narrowed composition space. Generally, it is desirable to find new perovskite compounds that can have high ORR activity from a high O *p*-band center and simultaneously exhibit good thermodynamic stability (i.e., have zero or low energy above the convex hull under ORR operating conditions). Based on the trends in O *p*-band center and stability in **Figure 4** and **Figure 5**, it is clear that as the O *p*-band center and thus ORR activity is increased, the stability simultaneously decreases. To further illustrate this trend, a plot showing the relationship between the calculated stability values and the O *p*-band center for all the data is contained in **Figure S4** in **Section 8** of the **SI**. Therefore, there is a tradeoff between maximizing both the activity and the stability of perovskite compounds to catalyze the ORR. Based on this tradeoff between ORR activity and material stability, it is easy to rationalize why a typical commercial SOFC cathode material like $La_{0.625}Sr_{0.375}Co_{0.25}Fe_{0.75}O_3$ (LSCF) has a lower activity but better stability than a state-of-the-art material like $Ba_{0.5}Sr_{0.5}Co_{0.75}Fe_{0.25}O_3$ (BSCF). It has been shown experimentally that BSCF exhibits very high activity but suffers from stability problems.[11, 13, 14] The *k** value of BSCF is approximately 200× higher than LSCF at T≈1000K, and our calculated stabilities of BSCF and LSCF are 124 meV/atom and 47 meV/atom, respectively. Overall, perovskites capable of exhibiting high activity and high stability will most likely contain a mixture of alkaline earth and rare earth elements on the A-site and a mixture of late transition metals and less redox-active elements on the B-site.

## 2.3. Phase stability under ORR operating conditions

The stability of the 2145 perovskite materials simulated in this work was analyzed under SOFC ORR operating conditions of $T$ = 1073 K, $p(O_2)$ = 0.21 atm, and $RH$ = 30% using the Pymatgen toolkit. The stability of each material was calculated as the energy difference above the convex hull, in meV/atom. As described in the screening procedure shown in **Figure 2**, we allow any high activity compounds that fall within 47 meV/atom above the convex hull to pass through our stability screening test, as it could warrant further investigation. We note here that our stability cutoff is most likely a conservatively low value that will increase false negatives (materials that we identify as not promising due to poor stability but are actually stable) while minimizing false positives (materials we identify as promising due to good stability but turn out to be unstable), as our goal is to find at least some promising materials rather than to be sure we have identified all of them in a particular composition space. In particular, this estimate is conservative as the DFT-based phase stability calculations performed here are only differences in total energies (which is approximately the differences in enthalpies) between materials, and ignores entropic terms that contribute to the overall free energy. While there are many factors that contribute to the overall entropy, such as vibrations, magnetic effects, and mixing, we know that the mixing entropy is significant and will drive stability of perovskite alloys with more than 3 components. In this study, we are concerned with stability at 1073 K, and for two sublattices at 50% mixing the entropy contribution to the free energy is $-2kT\left(\frac{1}{2}\ln\left(\frac{1}{2}\right)\right)$ per formula unit, which is about 64 meV/formula unit, or about 13 meV/atom, as there are 5 atoms per perovskite formula unit.

**Figure 6** contains plots of the predicted values of $k^*$ (from the O $p$-band calculations) as a function of the energy above the convex hull. **Figure 6A** shows data of all perovskite compounds simulated in this work. The symbol colors represent different families of perovskite compounds based on the number of alloying elements: ternary systems (red), quaternary systems (blue), and quinary systems (purple). Materials denoted with an "x" symbol are insulating, i.e., they fail elimination criterion 3 from **Figure 2**. Materials denoted with an "o" symbol pass elimination criterion 3. The green highlighted region of **Figure 6A** and **Figure 6B** is the region containing materials that have predicted $k^*$ values higher than LSCF (they pass elimination criterion 1) and have calculated stabilities within 47 meV/atom of the convex hull (they also pass elimination criterion 2). Also indicated on **Figure 6** are the approximate $k^*$ values for LSCF (log $k^* \approx$ -6.0 cm/s) and BSCF (log $k^* \approx$ -3.8 cm/s). Since LSCF is one of the best and most widely-implemented SOFC cathode materials and BSCF is one of the highest activity ORR materials known to date, they both represent significant milestones of performance. **Figure 6B** focuses on the region of **Figure 6A** where the stability is bounded by 100 meV/atom and the activity is bounded by log *k** of -12 cm/s to more clearly show the number of materials that lie in certain regions of this activity-versus-stability space. The data of **Figure 6B** clearly show that there are many compounds predicted to have surface exchange rates that exceed that of LSCF and stabilities better than BSCF (BSCF is 124 meV/atom above the convex hull), many of which might be of potential commercial interest. **Figure 6C** shows all materials contained within the green-highlighted portion of **Figure 6B**. It is clear from **Figure 6C** that while there are many materials that pass elimination criteria 1 and 2, many of these materials are insulating. **Figure 6D** further refines the data shown in **Figure 6C** by only displaying the materials that pass all elimination criteria. A summary of these promising SOFC cathode materials is provided in **Table 1** and the data spreadsheet as part of the **SI**.

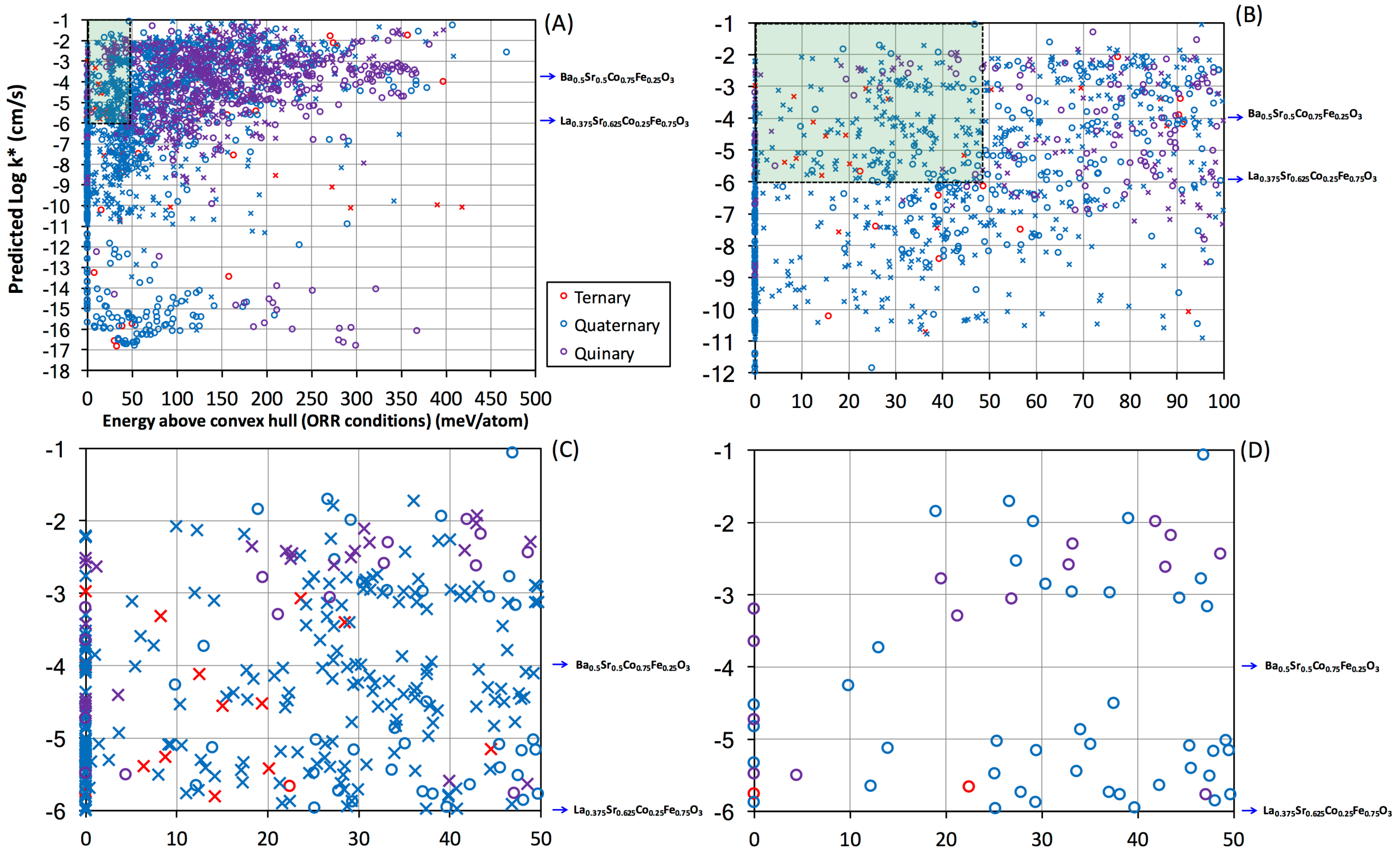

**Figure 6.** (A) Plot of predicted surface exchange coefficients log *k** as a function of stability under high temperature ORR conditions for all screened perovskites, given as the energy above the convex hull from Pymatgen phase stability analysis. Panel (B) focuses on a narrower range of log *k** and stability for increased visibility. The red, blue, and purple symbols represent perovskite

compositions that are ternary, quaternary and quinary alloys, respectively. The materials marked with "x" symbols are insulating, while those with "o" symbols are metallic. The predicted values of log $k^*$ for the common commercial cathode material $La_{0.625}Sr_{0.375}Co_{0.25}Fe_{0.75}O_3$ (LSCF) and state-of-the-art $Ba_{0.5}Sr_{0.5}Co_{0.75}Fe_{0.25}O_3$ (BSCF) are indicated by the blue arrows. Our materials screening has yielded many materials with predicted $k^*$ values higher than LSCF, some of which are stable under high temperature ORR conditions. This region of predicted high activity and stability within 47 meV/atom of the convex hull is highlighted in green. (C) Plot of the green highlighted portion from panels (A) and (B) showing materials that pass elimination criteria 1 and 2. (D) Same as panel (C), except only materials that pass all elimination criteria are shown. These materials are also listed in Table 1 and in the data spreadsheet as part of the SI.

**Table 1** contains the list of top 20 materials from **Figure 6D** that pass all elimination criteria, making this collection of materials the most physically relevant set of promising ORR materials for further investigation. The complete list of all materials which pass all screening criteria is provided in the data spreadsheet as part of the **SI**. Generally speaking, the promising materials in **Table 1** contain A-site compositions that are either entirely an alkaline earth element (particularly Ba), or a mixture of alkaline and rare earth elements. The B-sites of materials in **Table 1** are predominantly occupied by late transition metals such as Fe and Co. In addition, many of these top-performing materials also contain a redox-inactive element on the B-site, such as Zr, Hf, Ta, Nb, Y, Sc or Re. The inclusion of these redox-inactive cations provides increased stability without sacrificing the high catalytic activity. Consistent with these trends, there have been a number of recent studies that include redox-inactive dopants to stabilize highly active cathodes. For example, a recent study detailing the performance of $BaCo_{0.4}Fe_{0.4}Zr_{0.1}Y_{0.1}O_3$ as a cathode displayed peak power densities of nearly 1 W/cm$^2$ at T = 500 °C and good long term stability in a humid atmosphere.[65] Other studies found that doping redox-inactive elements such as Y,[66] Nb,[67] and Zr[68, 69] into $Ba_{0.5}Sr_{0.5}Co_{0.8}Fe_{0.2}O_3$ showed better resistance to degradation at high operating temperatures of about T>700 ˚C. As another example, experiments of Nb doping[45] in $SrCoO_3$ and Nb and Ta co-doped[70] $SrCoO_3$ displayed highly active cathodes with encouraging long-term stability, while doping of Nb and Fe into $SrCoO_3$, Sc and Nb into $SrCoO_3$ and Ag and Nb into $SrCoO_3$ all demonstrated very low ASR values.[71-73] These recent findings corroborate the general finding of this work that the inclusion of less redox active elements on the B-site can offer enhanced cathode stability without sacrificing high ORR activity.

As one would generally like to maximize both the activity and stability to create an optimal cathode material, we have shown that the computational screening techniques used here provide useful insight into developing schemes to not only discover new material compositions but also improve the activity and stability of known promising compounds. If one has a high activity perovskite alloyed purely from alkaline earth elements on the A-site and late transition metals on the B-site (a canonical example of which is BSCF), the material most likely becomes unstable over time. Improving the stability of a material like BSCF without dramatically reducing its activity can be accomplished by alloying the A-site with a small fraction of rare earth elements, and/or alloying the B-site with some redox-inactive elements such as Zr, Hf, Nb, Ta, Y, Sc or Re. The stability of BSCF (energy above hull = 124 meV/atom) relative to $Ba_{0.5}Sr_{0.5}Co_{0.625}Fe_{0.5}Zr_{0.125}O_3$ (BSCFZ),[68, 69] $Ba_{0.5}Sr_{0.5}Co_{0.625}Fe_{0.5}Nb_{0.125}O_3$ (BSCFN)[67] and $Ba_{0.5}Sr_{0.5}Co_{0.625}Fe_{0.5}Y_{0.125}O_3$ (BSCFY)[66] (energies above hull of 78, 65 and 101 meV/atom, respectively), as well as the stability of $BaFeO_3$ (BFO)[35] (energy above hull = 77 meV/atom) relative to $BaFe_{0.125}Zr_{0.875}O_3$ (BFZ)[61] (energy above hull = 0 meV/atom) and $BaFe_{0.875}Nb_{0.125}O_3$ (BFN)[43] (energy above hull = 69 meV/atom) represent concrete examples of this effect.

There has been a recent effort in the SOFC research community to minimize or eliminate the usage of alkaline earth elements in the cathode (Ca, Sr, or Ba) due to their tendency to segregate to the surface, react with the electrolyte, or simply result in cathode bulk instability. To help support these efforts, we have provided **Table S3** in **Section 10** of the **SI** that contains all materials in this work that pass our elimination criteria and do not contain alkaline earth elements.

**Table 1.** List of the top 20 most active materials which pass all elimination criteria. The material compositions, calculated O *p*-band center, predicted values of *k** based on O *p*-band calculations, and calculated stability under ORR conditions for materials which pass all elimination criteria are provided. Materials are listed in descending order by their predicted *k** values. As a point of comparison, the experimental values of log *k** for LSCF and BSCF are -6.0 cm/s and -3.8 cm/s, respectively. The complete list of all materials which pass all screening criteria is provided in the data spreadsheet as part of the SI.

| Material Composition | Calculated O *p*-band center (eV) | Predicted log k* (cm/s) | Calculated energy above convex hull (meV/atom) |
|---|---|---|---|
| $BaFe_{0.5}In_{0.5}O_3$* | -0.843 | -1.066 | 46.9 |
| $BaFe_{0.875}Al_{0.125}O_3$ | -1.105 | -1.943 | 39.0 |
| $Ba_{0.75}Sr_{0.25}Fe_{0.875}Ga_{0.125}O_3$ | -1.118 | -1.986 | 41.9 |
| $BaFe_{0.75}Y_{0.25}O_3$ | -1.119 | -1.990 | 29.1 |
| $Ba_{0.75}Sr_{0.25}Fe_{0.875}Al_{0.125}O_3$ | -1.176 | -2.180 | 43.4 |
| $Ba_{0.75}Sr_{0.25}Fe_{0.75}Y_{0.25}O_3$ | -1.212 | -2.301 | 33.2 |
| $BaFe_{0.75}Ta_{0.25}O_3$ | -1.282 | -2.535 | 27.3 |
| $Ba_{0.5}Sr_{0.5}Fe_{0.75}Sc_{0.25}O_3$ | -1.299 | -2.592 | 32.8 |
| $Ba_{0.5}Sr_{0.5}Fe_{0.875}Y_{0.125}O_3$ | -1.308 | -2.622 | 42.9 |
| $SrCo_{0.5}Sc_{0.5}O_3$ | -1.354 | -2.776 | 46.6 |
| $BaFe_{0.125}Co_{0.125}Zr_{0.75}O_3$ | -1.355 | -2.779 | 19.5 |
| $BaFe_{0.875}Re_{0.125}O_3$ | -1.378 | -2.856 | 30.4 |
| $SrFe_{0.875}Al_{0.125}O_3$** | -1.409 | -2.960 | 33.1 |
| $Ba_{0.5}Sr_{0.5}Fe_{0.875}Nb_{0.125}O_3$ | -1.439 | -3.060 | 26.8 |
| $Ba_{0.5}La_{0.125}Zn_{0.375}NiO_3$ | -1.479 | -3.194 | 0.0 |
| $Ba_{0.875}Sn_{0.125}Fe_{0.125}Zr_{0.875}O_3$ | -1.509 | -3.295 | 21.2 |
| $BaFe_{0.625}Mn_{0.25}Zr_{0.125}O_3$ | -1.615 | -3.649 | 0.0 |
| $BaFe_{0.75}Ru_{0.25}O_3$ | -1.640 | -3.733 | 12.9 |
| $SrCo_{0.75}Fe_{0.25}O_3$ | -1.706 | -3.954 | 44.5 |
| $BaFe_{0.5}Pt_{0.5}O_3$ | -1.799 | -4.265 | 9.8 |

*Lower In-doping levels were also promising, and these compositionally similar compounds are listed in the data spreadsheet provided in the **SI**.

** Ga and In were also promising dopants at this composition, and these compositionally similar compounds are listed in the data spreadsheet provided in the **SI**.

As described in **Section 1**, we also tested the stability of a subset of the promising materials listed in **Table 1** in the presence of competing Ruddlesden-Popper and hexagonal phases, as well as in the presence of common electrolyte materials. Full details of these analyses can be found in **Section 11** (stability with Ruddlesden-Popper and hexagonal phases) and **Section 12** (stability with common electrolytes) of the **SI**, respectively. Here, we provide a brief summary of these results. With regard to the stability of promising perovskites in the presence of competing phases, we found that the addition of the Ruddlesden-Popper and hexagonal phases did not destabilize the perovskite phase past our stability elimination criterion. However, in some cases the hexagonal phase was found to be nearly degenerate in energy with the perovskite phase, which may result in the presence of secondary phases during synthesis. Thus, careful control over synthesis conditions may be required to facilitate the formation of a single phase perovskite material. With regard to the stability of promising perovskites in the presence of common electrolyte materials such as yttria-stabilized zirconia (YSZ), gadolinium-doped ceria (GDC) and $Bi_2O_3$, we found that only $Ba_{0.5}La_{0.125}Zn_{0.375}NiO_3$ was compatible with these common electrolytes. However, we note that no kinetic considerations are included here, which might enable even unstable phases to coexist for long periods of time. In addition, minor compositional engineering and barrier layers can be used to reduce reactivity between the cathode and electrolyte.

Finally, we note here that from the standpoint of economic viability and sustainability, some of the compounds listed in **Table 1** may not be feasible to produce on a mass scale due either to their high cost and/or scarcity. In particular, compounds containing Re, Sc, Pt or Rh are not feasible candidates for large scale production as their precursor salts used to synthesize the perovskite phase are in the range of \$100-200 per gram. However, the other materials, which contain elements Zr, Ba, Y, Zn, Ta, Nb and In have approximate prices (from Sigma-Aldrich) ranging from less than \$1/gram (Ba, Zr) to a few dollars per gram (Y, Zn, Ta, Nb) to about \$9/gram (In). From the Sigma-Aldrich, one can purchase LSM and LSCF cathode materials for about \$5 and \$8 per gram, respectively. Therefore, the elements present in the promising materials found in this study that are not typically used in commercial SOFC cathode materials are reasonably abundant and affordable compared to commonly used cathode materials, thus making many of these promising cathode materials economically viable in large-scale SOFC production.

# 3. Summary and Conclusions

In this work, we have conducted high-throughput density functional theory-based screening of 2145 perovskites in search of stable and highly active ORR cathode materials, and have verified that our screening methodology reproduces the experimental activity, stability, and conduction properties of well-studied cathode materials. We used the bulk oxygen *p*-band center as an electronic structure descriptor for the surface exchange coefficient *k**, which is correlated to the overall ORR activity. We analyzed the thermodynamic phase stability under typical ORR operating conditions using the phase stability analysis tools contained in the Pymatgen toolkit. We systematically eliminated materials that do not pass each of our established elimination criteria: (1) material has a predicted *k** less than that of LSCF, (2) material has a calculated stability > 47 meV/atom (the stability value of LSCF) above the convex hull, and (3) material has a nonzero electronic bandgap or charge transfer gap. We found that of the 11 known promising cathode materials with measured *k** values greater than that of LSCF given in **Figure 3**, our *k** screening method would have found 9 of them had they not already been known. In addition, from the materials examined here, we have provided a list of promising new cathode materials that have, to our knowledge, not previously been studied, but passed all of our screening criteria and thus are expected to have high activity and stability under ORR operating conditions. Some of the top materials include the following: $Ba_{0.75}Sr_{0.25}Fe_{0.875}Ga_{0.125}O_3$, $SrCo_{0.5}Sc_{0.5}O_3$, $BaFe_{0.75}Ta_{0.25}O_3$, $BaFe_{0.125}Co_{0.125}Zr_{0.75}O_3$, $BaFe_{0.875}Re_{0.125}O_3$, and $Ba_{0.5}La_{0.125}Zn_{0.375}NiO_3$. We note here that of these top materials, the use of $Ba_{0.75}Sr_{0.25}Fe_{0.875}Ga_{0.125}O_3$, $BaFe_{0.125}Co_{0.125}Zr_{0.75}O_3$ and $Ba_{0.5}La_{0.125}Zn_{0.375}NiO_3$ is most likely as they contain constituent elements that are affordable and widely available. A complete summary of the promising materials found in this work is provided in **Table 1** and the data spreadsheet as part of the **SI**. We have separately verified that a selected subset of the top materials in **Table 1** are stable in the presence of competing Ruddlesden-Popper and hexagonal phases.

In addition to the discovery of new SOFC cathode materials to efficiently catalyze the ORR reaction, we have also used the O *p*-band center descriptor and stability analysis to examine the qualitative dependence of *k** and stability on the A- and B-site composition. Broadly, the inclusion of alkaline earth elements on the A-site will increase the value of the O *p*-band center and thus *k**, but will simultaneously tend to destabilize the material. The inclusion of late transition metal elements on the B-site, such as Fe, Co and Ni, will also tend to increase *k** and decrease stability. These findings corroborate known alloying trends from the experimental literature, where the highest measured values of *k** generally result from perovskites alloyed with these sets of elements. These results suggest one can increase the stability of highly active materials by alloying the A-site with a small fraction of rare earth elements, and/or alloying the B-site with some redox-inactive elements such as Zr, Hf, Nb, Ta, Y, Sc or Re. Such approaches have already been demonstrated in the literature and we discussed some examples in **Section 2.3**.

Overall, the computational screening schemes used here have been instrumental in discovering new perovskite compounds and suggesting methods to improve existing compounds. Experimental testing of the promising materials provided here and further refinement of our computational screening model (for instance, to incorporate thermal expansion coefficient mismatch between cathode and electrolyte, other cation ordering schemes, more realistic O non-stoichiometry, etc.) are promising next steps to improve the effectiveness of our approach for the discovery and implementation of new stable, highly active ORR catalysts for the next generation of SOFCs. In addition to their use in high temperature SOFCs, the materials discovered in this work may be useful in other related applications that can utilize oxygen exchange, such as low temperature OER/ORR,[11, 12, 74] $CO_2$ capture via chemical looping,[75-77] and thermochemical water splitting,[78-81] although additional research is required to thoroughly examine the characteristics of these promising perovskites for these emergent energy applications.

# 4. Computational Methods

All total energy and electronic structure calculations in this work were performed using Density Functional Theory (DFT) as implemented by the Vienna *Ab Initio* Simulation Package (VASP).[82] The DFT calculations were automated in a high-throughput manner by using the Materials Simulation Toolkit (MAST), which interfaces directly with VASP.[16] A planewave basis set was used to represent the electron wavefunctions, and the planewave cutoff energy was set to be 30% larger than the highest pseudopotential planewave cutoff energy. The Perdew-Burke-Ernzerhof (PBE)-type pseudopotentials[83] utilizing the projector augmented wave (PAW)[84] method were used to represent each element type. In general, all specific pseudopotentials used are the same as those used within Pymatgen and the Materials Project, so that consistent calculations of phase stability were obtained.[64] For all calculations, the generalized gradient approximation (GGA) was used as the exchange and correlation functional. For materials containing transition metals, the Hubbard $U$ correction method (GGA+$U$)[85] was implemented with effective $U$ values equal to those used in the Pymatgen package and Materials Project.[17, 86] The $U$ values are tabulated in **Table S1** of **Section 5** of the **SI**. The Monkhorst-Pack scheme was used for the reciprocal space integration of the Brillouin zone for all materials.[87] Reciprocal space k-point meshes of 4×4×4, 2×2×2, and 4×4×2 were used for all perovskite (40 atoms/cell), hexagonal (120 atoms/cell), and Ruddlesden-Popper (56 atoms/cell) material phases, respectively. (See **Figure 1** for more information on these structures and **Section 3** of the **SI**). For all calculations, the choices of k-point mesh and planewave cutoff energy result in total energy convergence of approximately 3 meV/supercell, and all calculations were done with spin polarization enabled. All materials were simulated as fully oxidized (i.e., with chemical formula $ABO_3$, with no oxygen vacancies), unless otherwise indicated. We note here that many, if not most of the materials considered here will exhibit some degree of off-stoichiometry via creation of oxygen vacancies. Investigating what the precise oxygen content for every perovskite considered here is beyond the scope of the current work, however a more detailed examination of the expected oxygen stoichiometry and its effect on the O $p$-band center and thus predicted activity and stability may be a topic for future study of the most promising compounds discovered here.

The O $p$-band center was calculated as the centroid of the densities of states (DOS) projected onto the 2$p$ orbitals of the O atoms using the following equation:

$$\bar{O}_{2p} = \frac{\int_{-\infty}^{\infty} E \cdot D_{O_{2p}}(E)dE}{\int_{-\infty}^{\infty} D_{O_{2p}}(E)dE} - E_{Fermi} \, , \tag{1}$$

where $\bar{O}_{2p}$ is the O $p$-band center, $E$ is the electron energy, $D_{O_{2p}}(E)$ is the DOS projected onto the 2$p$ orbitals of O, and the integrals are taken over all states, not just filled states. All calculated values of the O $p$-band center are given with respect to the Fermi energy $E_{Fermi}$. We note here that it is also reasonable to calculate the O $p$-band center by integrating over only the filled O states, as surface exchange involves the O exchanging electrons with the filled O states. However, we found that slightly improved trends were obtained by integrating over all O states. This may be because the empty O states provide additional electronic structure information (such as changing O $p$-band-B *d*-band hybridization with composition changes), thus providing a better overall correlation when examining a large composition space of perovskite materials.

All phase stability calculations were conducted using the phase diagram analysis tools contained within Pymatgen. Additionally, the chemical potentials of $O_2$ and $H_2$ were shifted to coincide with values indicative of high temperature SOFC ORR operating conditions of $T$ = 800 °C (1073 K), $p(O_2)$ = 0.21 atm and relative humidity (*RH*) of 30%. Additional details of these calculations can be found in **Section 2** and **Sections 5-7** of the **SI**.

The electronic bandgap and charge transfer gap were calculated using the (projected) densities of states for each material. The bandgap is the energy difference between the highest filled electronic state (i.e., the Fermi level in DFT calculations) and the lowest unoccupied energy state. The definition of the charge transfer gap differs slightly from the electronic bandgap. Following the work of Ref. [88], we have calculated the charge transfer gap to be the energy difference between the highest filled O 2*p* state and the lowest unoccupied O 2*p* state. A small value of the charge transfer gap between occupied and unoccupied O states is understood to more efficiently facilitate charge transfer between O atoms reacting at the perovskite surface during catalysis (i.e., the material is a better conductor).[88, 89]

# Acknowledgements

This work was supported by the US Air Force Office of Scientific Research through grants No. FA9550-08-0052 and No. FA9550-11-0299. Funding for Dane Morgan and 50% of Ryan Jacobs for this work was provided by the NSF Software Infrastructure for Sustained Innovation (SI2) award No. 1148011. Computational support was provided by the Extreme Science and Engineering Discovery Environment (XSEDE), which is supported by National Science Foundation Grant No. OCI-1053575. This research was also performed using the compute resources and assistance of the UW-Madison Center For High Throughput Computing (CHTC) in the Department of Computer Sciences.

# Supporting Information:

## Materials discovery and design principles for stable, high activity perovskite cathodes for solid oxide fuel cells

Ryan Jacobs,[1] Tam Mayeshiba,[1] John Booske[2], and Dane Morgan[1,*]

[1]Department of Materials Science and Engineering, University of Wisconsin- Madison, Madison, WI.

[2]Department of Electrical and Computer Engineering, University of Wisconsin- Madison, Madison, WI.

*Corresponding author e-mail: ddmorgan@wisc.edu

**Summary of contents of this SI:**

## 1. Complete data list of the materials examined in this work

A collection of the data obtained for all materials simulated in this work is available online in spreadsheet form as part of this **SI**. The spreadsheet contains calculated data of the energy above the convex hull under ORR operating conditions ($T$ = 1073 K, $p(O_2)$ = 0.21 atm, $RH$ = 30%), the calculated O $p$-band center and predicted value of $k^*$ using the trends discussed in the main text, and the band gap and charge transfer gap values for each material. For a brief discussion of how to obtain the chemical potentials of $O_2$ and $H_2$ used in this work, see **Section 7** of this **SI**. In the data spreadsheet, material compositions are listed as a sequence of elements and numbers, where the numbers indicate the number of atoms of the particular element used in the simulated supercells. For example, the entry "Ba1Sr7V8O24" is the compound $Ba_{0.125}Sr_{0.875}VO_3$. We used 40-atom 2×2×2 perovskite supercells for all screening calculations; therefore, there are 8 $ABO_3$ formula units per supercell (i.e., $A_8B_8O_{24}$), and the compound "Ba1Sr7V8O24" is a compound with 1/8 Ba and 7/8 Sr on the A-site, and pure V on the B-site.

## 2. Material databases and phase stability calculations

The database of calculated materials in the Materials Project consists of most of the entries from the Inorganic Crystal Structure Database (ICSD), and also contains structures from other repositories or structures simulated by various research groups. As of 2017, approximately 70,000 inorganic compounds have been calculated and tabulated in the Materials Project database.[1] Many of the base perovskite compounds that are part of our undoped ternary structures (e.g. $LaCrO_3$, $PrNiO_3$, etc.) are contained in these databases. However, the doped quaternary (e.g. $La_{0.75}Sr_{0.25}MnO_3$) and quinary (e.g. $Ba_{0.5}Sr_{0.5}Co_{0.75}Fe_{0.25}O_3$) perovskites, as well as most hexagonal and Ruddlesden-Popper phases, are not present in the Materials Project at the time of this writing. For oxide materials, the Materials Project database consists of DFT-calculated energies for each material. These DFT energies are all at conditions of $T$ = 0 K, but have been shifted using known corrections for the gaseous $O_2$ species. In this way, the DFT-calculated $T$ = 0 K solid phase energies are shifted using the temperature and pressure dependence of the gas species, and are an approximation for standard conditions of $T$ = 298 K and $P$ = 1 atm. Additional details of how this energy shift is used for oxide materials is discussed in **Section 5** of this **SI**.

The high temperature chemical phase stability of all compounds screened in this study was analyzed using the multicomponent phase diagram modules contained within the Pymatgen toolkit (version 4.2.0).[2] More information about how to use Pymatgen to conduct multicomponent phase stability analysis is provided in **Section 6** of this **SI**. It was assumed that every potential high temperature SOFC cathode material would be subject to an environment that is open both to $O_2$ and $H_2$, consistent with possible reactions with oxygen and water vapor. The chemical potential of $O_2$ was set such that the temperature was equal to 800 °C (1073 K) and the partial pressure of $O_2$ was equal to 0.21 atm. The chemical potential of $H_2$ is set by the chemical potential of $O_2$ and equilibrium with $H_2O$ vapor at 1073 K. We assumed a humid operating environment with a relative humidity (*RH*) of 30%, which is an approximate value for the amount of $H_2O$ present in ambient air which is often the source of $O_2$ gas for high temperature SOFC operation.[3, 4] More information on how these chemical potentials were calculated can be found in **Section 7** of this **SI**.

For all phase stability calculations, the calculated DFT total energies for each compound must be shifted to account for various corrections used within the Pymatgen framework. These shifts include (1) the gas phase $O_2$ shift and DFT overbinding corrections (which are applied to the solid phase energies),[5] and (2) the solid phase energy shift needed when phase stability calculations consist of some materials modeled by GGA and others by GGA+$U$ (e.g. a GGA calculation of metallic Fe and GGA+$U$ calculation of $LaFeO_3$ in the La-Fe-O system).[6] More information on the details of these energy shifts can be found in the Materials Project online documentation and in **Section 5** of this **SI**.[7]

## 3. Material structures explored

Perovskites that possess rhombohedral (space group $R\bar{3}c$), orthorhombic (space group $Pbnm$), or cubic ($Pm\bar{3}m$) symmetries in the ground state were all scaled in size to construct 2×2×2 supercells (40 atoms/cell). In all cases, the structures were fully relaxed (volume + ions). For select perovskite materials that were predicted to have high surface exchange coefficient $k^*$ (via a high calculated O $p$-band center, see **Section 2.1**), hexagonal (space group $P6_3cm$) and Ruddlesden-Popper (general formula $A_2A'_{n-1}B_nO_{3n+1}$ with n = 1, space group $I4/mmm$) variants were also modeled to compare the stability of these competing phases with the perovskite phase. Whenever possible, the structure for the composition of interest was obtained from the Materials Project database.[1] For compositions not tabulated on the Materials Project, the structure of a chemically similar perovskite was used as the starting structure. For example, if the material $ANiO_3$ had a structure tabulated on Materials Project, but $A'NiO_3$ did not, the structure of $ANiO_3$ was used as the starting point for simulating $A'NiO_3$. This method bias getting the correct tilt systems, but doesn't guarantee it. Missing a correct tilt system may introduce some error in the O $p$-band and stability calculations, but we believe this error is acceptable in order to simulate and analyze a large number of systems. In the future, specific promising systems subject to more detailed study will require a more in-depth analysis of the impact of specific tilt system on stability and O *p*-band center. Simulating the hexagonal and Ruddlesden-Popper competing phases allowed us to determine if they were unstable, able to coexist with the perovskite phase, or potentially destabilize the perovskite phase (see **Section 11** of this **SI**). As shown in **Figure 1**, when multiple cations are considered on the A- or B-sites the cations were equally distributed among all available A- and B-site containing [001] ([0001] for hexagonal structures) planes in the simulated supercells (i.e., every plane normal to [001] direction ([0001] direction for hexagonal structures) was given the same composition), and compounds containing multiple cations on the A- and/or B-site were ordered along the [001] direction. For example, all compounds with composition $A_{0.75}A'_{0.25}BO_3$ had the same ordering of A' cations along the [001] direction. We note here that the precise ordering of the cations affects the value of the O *p*-band center. A recent study examining the surface exchange in $La_{0.6}Sr_{0.4}CoO_3$ as a function of strain demonstrated that the ordering of Sr on the A-site lattice resulted in an O *p*-band center change between about 0.05-0.1 eV.[8] Based on our trend of surface exchange versus O *p*-band center in **Figure 3**, a 0.1 eV change in the O *p*-band center corresponds to about a 0.33 cm/s change in log *k**.

## 4. Generation of perovskite compositions and materials screening

In this study, we considered perovskite materials that represent ternary, quaternary, and quinary compounds. A comprehensive list of all materials investigated in this work is provided in of the data spreadsheet as part of this **SI**. We began our study with an initial set of ternary and quaternary compounds (described in more detail below). Analysis of the predicted activity and calculated stability of this initial set of compounds (see **Section 2** of the main text for all results and discussion) provided guidance for subsequent materials alloying. In this way, we produced additional sets of quaternary and quinary compounds to best explore the relevant composition space which we predicted to contain perovskite compounds exhibiting the highest stability and predicted activity.

**Ternary compounds:**

We began our perovskite materials screening with the rare earth and alkaline earth elements (La, Y, Pr, Ca, Sr, Ba) comprising 100% of the A-site and the 3*d* transition metal row (Sc, Ti, V, Cr, Mn, Fe, Co, Ni, plus Ga) comprising 100% of the B-site, e.g. $YVO_3$, $LaFeO_3$, etc. This database had 54 compounds. Later in the study, additional miscellaneous ternary compounds were studied based on our continually evolving findings of which elements maximize the stability and activity of perovskites. These compounds contained elements such as Al, Zr, Hf, and Y on the A-site. In all, there were 72 ternary compounds studied in this work. This initial database of materials was motivated by observing which elements typically comprise perovskite materials over a range of technologically important applications, such as: catalysis and solid oxide fuel cells,[9-12] oxide electronics,[13-15] transistor dielectrics,[16, 17] magnetic tunnel junctions[18, 19] and solid state memory.[20, 21] The ternary perovskites considered here have A-site and B-site oxidation states (A+/B+) of either the 3+/3+ or 2+/4+ variety. The state of 3+/3+ is attained by selecting a rare earth cation (nominally 3+ oxidation state) on the A-site and 2+/4+ occurs when the A-site is an alkaline earth (nominally 2+ oxidation state) cation. In this way, the B-site cation oxidation state changes from 3+ (rare earth on A-site) to 4+ (alkaline earth on A-site). It is known that some of the ternary materials considered in this initial database are not stable as the perovskite structure, such as $BaNiO_3$, $CaMnO_3$, etc., but they were included to extract trends.

**Quaternary compounds:**

The ternary perovskite compounds described above served as a basis set of materials for separate alloying of the A-site and B-site to generate a large set of quaternary compounds. Where applicable, A-site dopants include 25 and 50% site fraction of the rare earth elements La, Y, Pr, Dy, Gd, Ho and Sm and alkaline earth elements Ca, Sr, and Ba; B-site dopants include 12.5, 25 and 50% site fraction of the elements Cr, Mn, Fe, Co, Ni and Mg. The designation of "where applicable" means, for instance, no Pr doping on the A-site was considered for $PrBO_3$-based materials. Based on results of the O *p*-band center for these preliminary materials (see **Section 2.1** of the main text), it was found that the (Ca, Sr, Ba)(Mn, Fe, Co, Ni)$O_3$ ternary materials yielded high predicted ORR activities. As a result, three additional subsets of quaternary materials were created. First, the A(Mn, Fe, Co, Ni)$O_3$ (A = alkaline earth Ca, Sr, or Ba) family of materials was further alloyed with 12.5, 25 and 50% B-site fraction of Mn, Fe, Co, Ni (where applicable) to ascertain if mixed B-site alloying of transition metal elements could further improve the predicted activity. Second, $BaFeO_3$ was used as a parent compound to further screen A-site concentrations of 12.5, 25, 50, 75, and 100% using the elements (Ca, Ce, Dy, Gd, Ho, La, Mg, Nd, Pr, Sm, Sr, Bi, Cd, and Sn) and B-site concentrations of 12.5, 25, 50, and 100% using the elements (Al, Cu, Ga, Hf, Ir, Mo, Nb, Os, Pd, Pt, Re, Rh, Ru, Ta, Tc, W, Y, Zn, Zr) to ascertain if these additional elements not previously considered could further increase the predicted activity and provide additional stability to the material. Third, we generated additional quaternary compounds belonging to the family Ba(Fe, Co, Ni)$O_3$ that were doped with 50, 75, and 87.5% B-site fraction of redox-inactive elements Hf, Nb, Zr. Later in the study, additional miscellaneous quaternary compounds were studied based on our continually evolving findings of which elements maximize the stability and activity of perovskites. The variety of compositions is too numerous to describe here, and instead the interested reader may refer to the data spreadsheet which is part of the **SI**. In total, the quaternary alloying schemes described here generated a pool of 1359 quaternary perovskite materials.

**Quinary compounds:**

As discussed regarding the generation of our quaternary set of materials and initial results in described in **Section 2.1**, the set of ternary materials belonging to (Ca, Sr, Ba)(Mn, Fe, Co, Ni)$O_3$ exhibit high predicted activity. We generated a set of quinary compounds by alloying the B-site of these ternary materials with 12.5, 25, and 50% site fraction Mn, Fe, Co, Ni, and, for each of these quaternary materials, further alloying the A-site with 50% of La, Y, Pr, Ca, Sr, and Ba, where applicable. Next, based on additional data of dopant elements which act to create stable compounds that also have high predicted activity, we generated additional quinary compounds belonging to the family (Pr, Ba)(Fe, Co, Ni)$O_3$ that are doped with 50, 75, and 87.5% B-site fraction of redox-inactive elements Hf, Nb, Zr. Finally, small subsets of additional quinary materials were generated over the course of this study to test different alloying elements based on promising parent compounds. These additional subsets are too numerous to list here, and are instead included in the list of all materials investigated in this study in the data spreadsheet which is part of the **SI**. In total, the quinary alloying schemes described here generated a pool of 714 quinary perovskite materials.

## 5. Energy shifts used in phase stability analysis

In order to use the Pymatgen toolkit for phase stability analysis, two energy shifts must be employed. In the next paragraph, we will first discuss the $O_2$ gas correction, followed by a discussion of the shifts needed when mixing GGA and GGA+*U* calculations.

The $O_2$ gas shift is necessary as DFT methods tend to overbind the $O_2$ molecule. As a result, the formation energies for oxides tend to be higher (i.e., more positive, less stable) than experimental values. Previous work has studied this $O_2$ gas shift in detail.[5, 10, 22] In the present work, we used the shift employed by the Materials Project so that our shifted energies were consistent with this database. This energy shift is 0.7023 eV/O, and was subtracted from the DFT calculated solid phase energy. For example, if the DFT calculated energy of $LaScO_3$ is –43.933 eV/formula unit, the shifted energy would be –43.933 eV/formula unit – (3 O/formula unit)(0.7023 eV/O) = -46.040 eV/formula unit. It is important to note that this $O_2$ gas shift is dependent on the exchange-correlation functional and pseudopotentials used. In this study the PBE functional and PAW-PBE type pseudopotentials were used, consistent with the Materials Project database.

The second shift needed to conduct phase stability analysis is only relevant for GGA+*U* calculations. This shift is required because the phase stability of compounds modeled with GGA+*U* will also draw upon materials modeled solely with GGA. For instance, in the La-Fe-O system, metallic Fe is modeled with GGA, while the $LaFeO_3$ perovskite is modeled with GGA+*U*. The details of these mixed GGA/GGA+*U* calculations are given in Ref. [6]. These energy shifts are subtracted from the calculated DFT energies. For example, if $LaFeO_3$ has a DFT calculated energy of -38.183 eV/formula unit, then the shifted energy, accounting for both the $O_2$ gas shift and the GGA/GGA+*U* mixing shift, would be -38.183 eV/formula unit – (1 Fe/formula unit)(2.733 eV/Fe) – (3 O/formula unit)(0.7023 eV/O) = -43.023 eV/formula unit. This GGA/GGA+*U* mixing shift is dependent on the transition metal element and the *U* value used (and also likely the exchange-correlation functional and pseudopotentials used). All *U* values and GGA/GGA+*U* mixing shifts used in this study are listed below in

**Table** S2, and they are the same values used in the Materials Project database.

**Table S2.** Collection of *U* values and GGA/GGA+*U* energy shifts used in this work.

| Element | *U* (eV) (*J* = 0 eV for all) | GGA/GGA+*U* shift (eV/ atom) |
|---|---|---|
| V | 3.25 | 1.682 |
| Cr | 3.7 | 2.013 |
| Mn | 3.9 | 1.681 |
| Fe | 5.3 | 2.733 |
| Co | 3.32 | 1.874 |
| Ni | 6.2 | 2.164 |

## 6. Using Pymatgen to conduct phase stability analysis of open systems

The Python Materials Genomics (Pymatgen) toolkit is a set of materials analysis tools written in the Python programming language.[2] For the stability analysis in this work Pymatgen version 4.2.0 was used. To conduct the stability analysis, we used the *PhaseDiagram* class within the *pymatgen.phasediagram.maker* module. The *PhaseDiagram* class requires information on the elements present, material composition, and (optionally) the chemical potential of gaseous reacting species if one is interested in environmental conditions different from DFT conditions. The information of material composition and energy are used in the *PDEntry* class from the *pymatgen.phasediagram.entries* module. Properly representing the data of material composition in the *PDEntry* class requires the use of the *Composition* class within the *pymatgen.core.composition* module. The energies of the materials used in the phase diagram analysis are the calculated DFT energies which have been appropriately shifted using the O energy shift and GGA/GGA+*U* mixing shifts described in **Section 5** of this **SI**. Once the phase diagram for the system of interest is created, the *PDAnalyzer* class in the *pymatgen.phasediagram.analyzer* module is used to compute the energy above the convex hull at the relevant composition in the phase diagram.

It is worth noting here that if one simply uses the default chemical potential values for $O_2$ and $H_2$, these chemical potentials are equal to their DFT-calculated values. Thus, the thermodynamic conditions under consideration for the phase stability analysis correspond to DFT conditions of T = 0 K and P = 0 atm but with energy shifts applied to all compounds (as discussed in **Section 5** of this **SI**) which corresponds approximately to standard conditions. These conditions are more oxidizing than typical high temperature ORR operating conditions of solid oxide fuel cells, thus we manually removed the calculated $O_2$ and $H_2$ entries from Pymatgen and substituted our calculated $O_2$ and $H_2$ chemical potentials indicative of ORR operating conditions. More information on how these chemical potentials were calculated is given in **Section 7** of this **SI**.

## 7. Chemical potentials used in phase stability analysis open to $O_2$ and $H_2$

The values of the chemical potentials for $O_2$ and $H_2$ used in the phase diagram analysis tools in Pymatgen were derived from standard gas phase thermodynamics equations. As all solid phase DFT energies are under conditions of $T = 0$ K and $P = 0$ atm, the temperature and pressure values typical of ORR operating conditions are built into the $O_2$ and $H_2$ gas chemical potentials. We calculated the chemical potential of $O_2$ using experimental data from the NIST chemistry webbooks[23] and standard thermochemistry equations as detailed in other works.[5, 10, 24-26] From our calculations, we obtained an O chemical potential of –6.25 eV/O at $T = 1073$ K and $p(O_2) =$ 0.21 atm, where this value is referenced to the enthalpy of $O_2$ gas at standard temperature and pressure. We calculated the H chemical potential using the value of the O chemical potential and equilibrium with water vapor. As with the case of $O_2$ gas, the NIST chemistry webbook is used to obtain experimental data of the free energy of $H_2O$. We assumed a typical humid environment with relative humidity of 30%. Using these values, we obtained an H chemical potential of -3.65 eV/H, where this value is referenced to the enthalpy of $H_2$ gas at standard temperature and pressure

## 8. Plots of additional materials trends

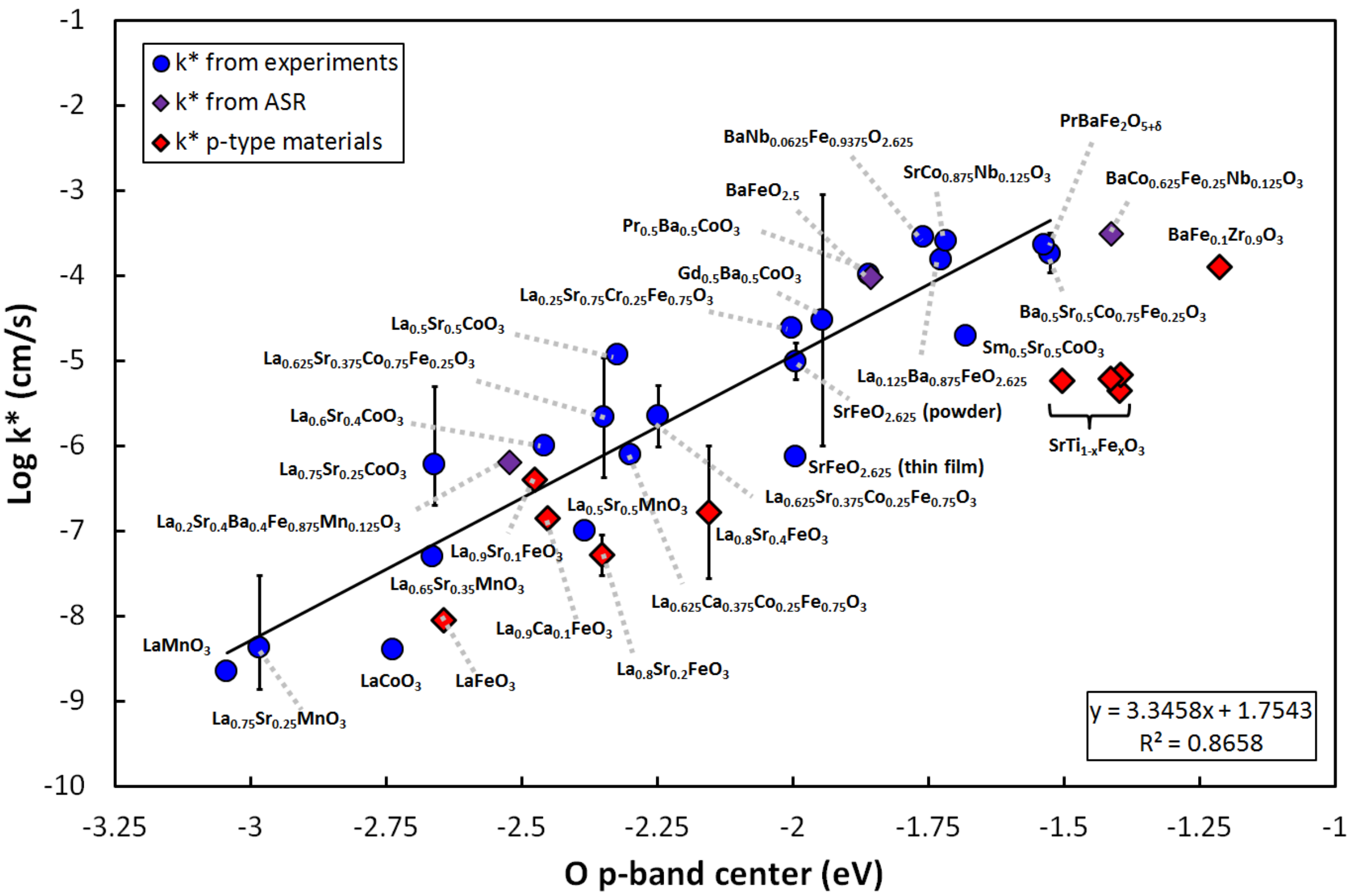

**Figure S1.** Plot of surface exchange coefficient log $k^*$ (in cm/s) as a function of calculated O $p$-band center (in eV). The O $p$-band center values are given relative to the Fermi energy for each material. The blue and red symbols are experimental surface exchange coefficient data. The line of best fit was made from the materials comprising the blue symbols, which were obtained from the collected values in Ref. [11], as well as the following materials and references: $La_{0.6}Sr_{0.4}CoO_3$: Ref. [27], $La_{0.5}Sr_{0.5}CoO_3$: Refs. [28, 29], $La_{0.25}Sr_{0.75}Cr_{0.25}Fe_{0.75}O_3$ (LSCrF): Ref. [30], $SrFeO_{2.625}$ (SF): Ref. [30] (powder), Ref. [31] (thin film), $La_{0.0625}Ba_{0.9375}FeO_{2.625}$ (LBF): Ref. [32], $BaNb_{0.0625}Fe_{0.9375}O_{2.625}$ (BFN): Ref. [33], $PrBaFe_2O_{5+d}$ (PBFO): Ref. [34], and $SrCo_{0.875}Nb_{0.125}O_3$ (SCN): Ref. [35]. The $k^*$ data for the purple symbols were obtained by using ASR data reported for $La_{0.2}Sr_{0.4}Ba_{0.4}Fe_{0.875}Mn_{0.125}O_3$ (LSBFMO): Ref. [36], $BaFeO_{2.5}$ (BF): Ref. [37], and $BaCo_{0.625}Fe_{0.25}Nb_{0.125}O_3$ (BCFN): Ref. [38] and the relationship of $k^*$ versus ASR as shown in **Figure S3** of this **SI**. The red symbols were plotted using experimental $k^*$ values for materials that were reported to be $p$-type conductors, which were obtained from the following references: $LaFeO_3$ and $La_{0.5}Sr_{0.5}FeO_3$: Ref. [39], $SrTi_{1-x}Fe_xO_3$: Ref. [31], $BaFe_{0.125}Co_{0.125}Zr_{0.75}Y_{0.125}O_3$: Ref. [40], and $BaFe_{0.1}Zr_{0.9}O_3$: Refs. [41, 42]. As shown here, the $p$-type materials do not fall very close to the line of best fit. This discrepancy is expected because, as explained in the main text, surface exchange in these materials is likely to be at least partially limited by electronic charge transfer, which is not expected to be well-correlated to the O $p$-band. The error bars were made by taking the maximum and minimum reported $k^*$ values (the fitted data points are the average $k^*$ values) from multiple references: $La_{0.8}Sr_{0.2}FeO_3$: Refs. [43, 44], $La_{0.6}Sr_{0.4}FeO_3$: Refs. [39, 45], and the references for the remaining materials are the same as provided in the **Figure 3** caption of the main text.

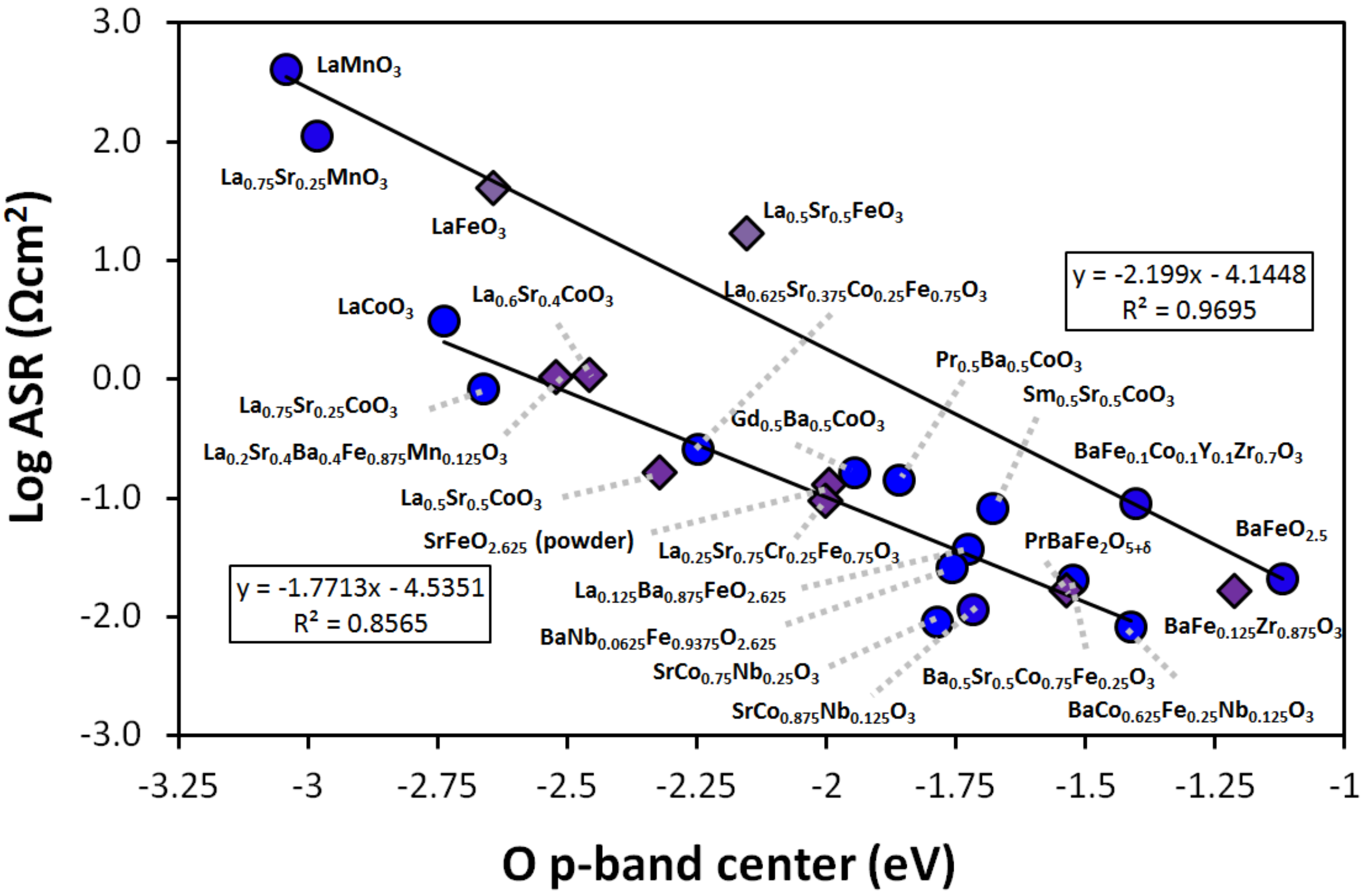

**Figure S2.** Plot of area specific resistance (ASR, in Ω-cm$^2$) as a function of calculated O *p*-band center (in eV). The O *p*-band center values are given relative to the Fermi energy for each material. The blue symbols are experimental ASR data, and the purple symbols were obtained by predicting ASR based on experimental *k** data (see **Figure S3** for the ASR versus *k** relationship). The lines of best fit were made from the materials comprising the blue symbols, which were obtained from the same references as listed in the caption of **Figure S1**. The bottom line of best fit is the group of materials where O transport is expected to be bulk-limited, and the top line of best fit is the group of materials where O transport is expected to be surface-limited (which are $LaMnO_3$, $La_{0.75}Sr_{0.25}MnO_3$, $BaFe_{0.1}Co_{0.1}Y_{0.1}Zr_{0.7}O_3$ and $BaFeO_{2.5}$).

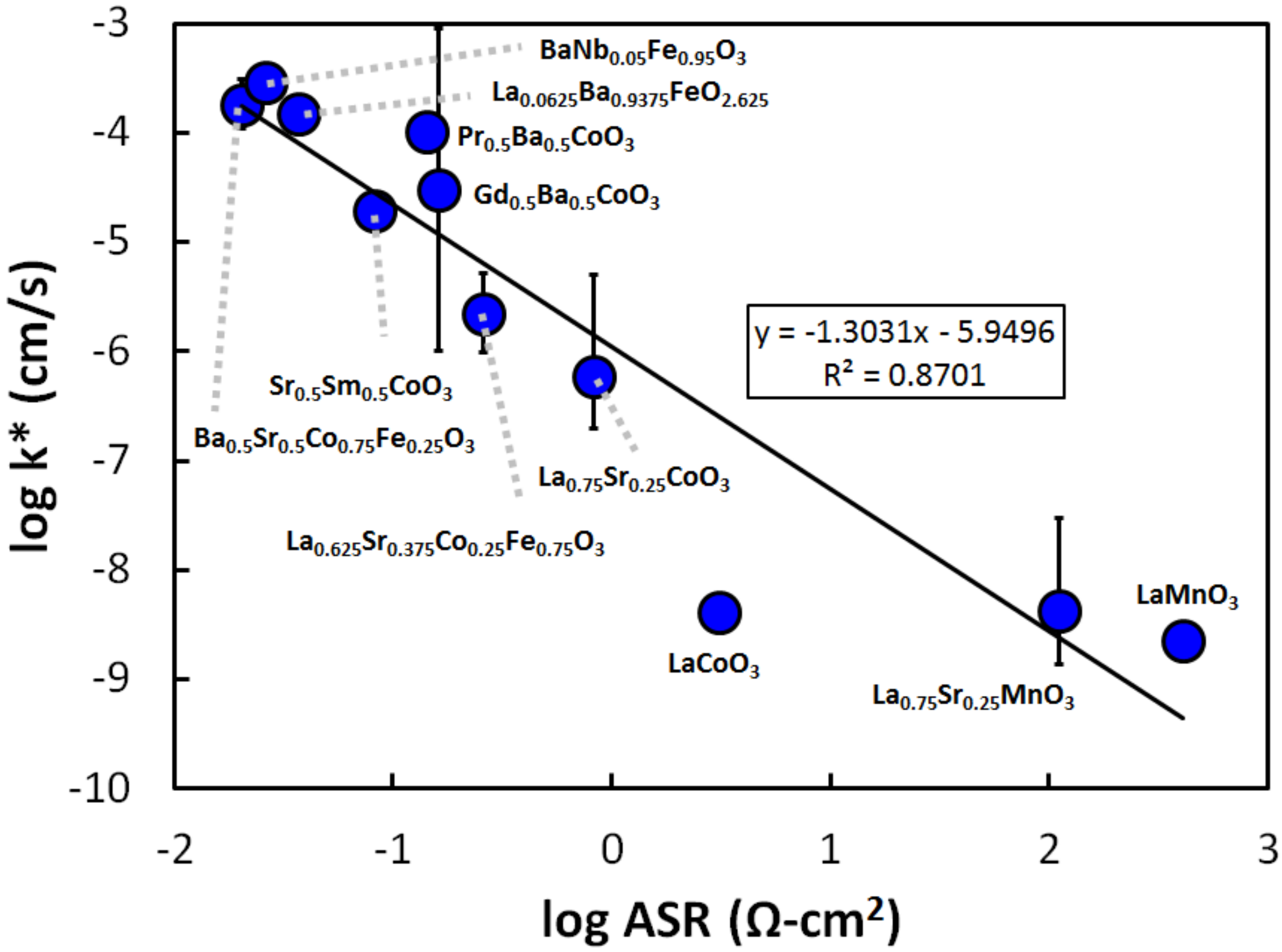

**Figure S3.** Plot of experimental $k^*$ versus experimental ASR for the set of materials from Ref. [11], except for $BaNb_{0.05}Fe_{0.95}O_3$ and $La_{0.0625}Ba_{0.9375}FeO_{2.625}$, which were obtained from Ref. [33] and Ref. [32], respectively. The linear relationship between $k^*$ and ASR was used to obtain predicted $k^*$ values in **Figure 3** of the main text and **Figure S1**, as well as predicted ASR values in **Figure S2**.

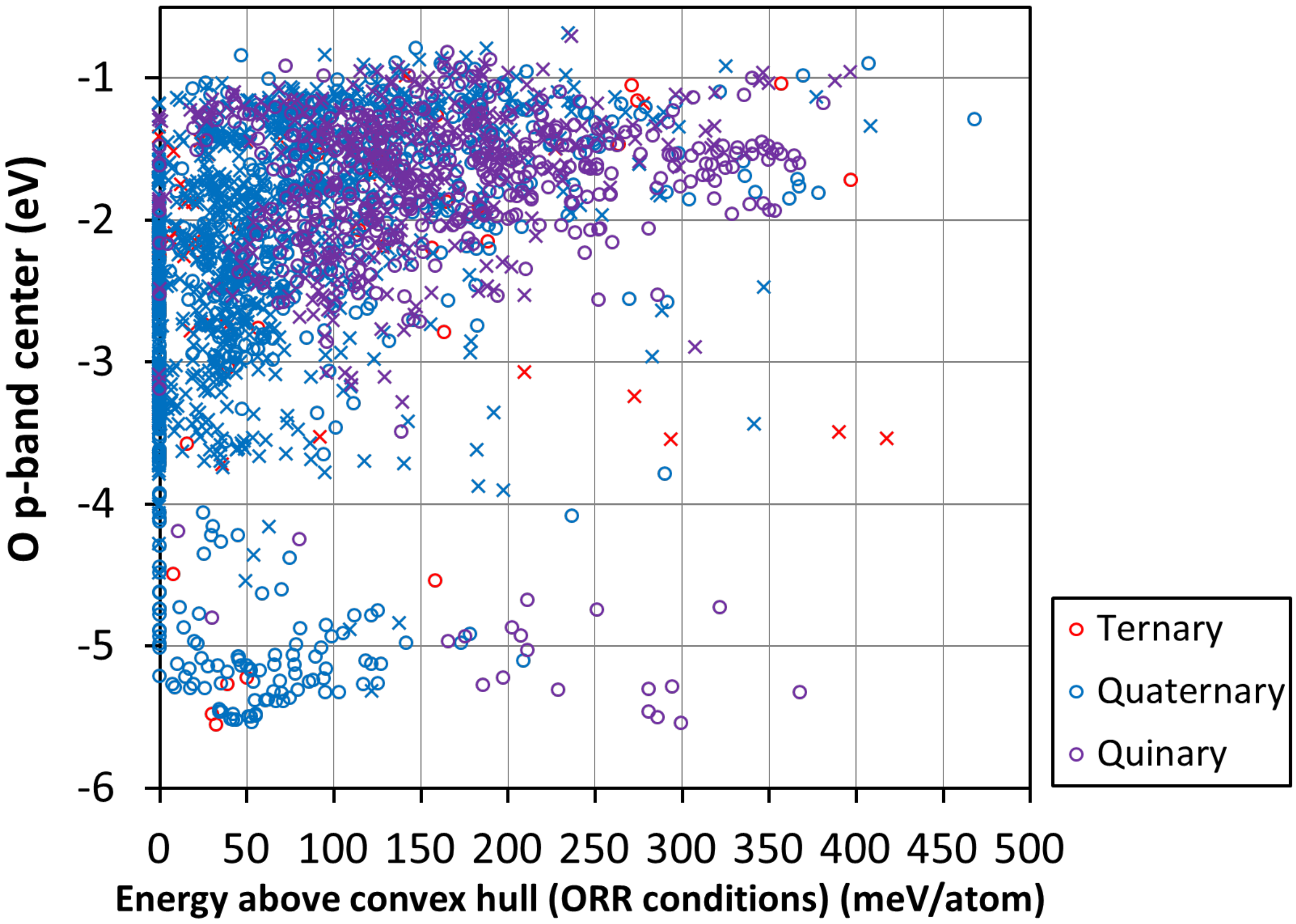

**Figure S4.** Plot of the calculated O *p*-band center as a function of stability as energy above the convex hull for all materials considered in this study. The symbol definitions are the same as **Figure 6** in the main text.

## 9. Model validation of well-studied compounds

**Table S2** contains data of experimental and predicted *k** and ASR data for well-studied cathode materials for validation of our model. Overall, the differences between the predicted and experimental ASR and *k** values are accurate to within about a factor of 5, although it should be noted that these materials are used in fitting the correlation used to predict *k**. In addition, our stability calculations correctly identify LSM, LSC and LSCF as stable compounds within our cutoff, and BSCF as an unstable compound. Note that the designation of LSCF as stable was by-design and is the stability cutoff as shown in **Figure 2** of the main text. Our simulations also predict zero bandgap and charge transfer gap for all materials (except the charge transfer gap of LSM), which is in agreement with the qualitative observation that the conductivity of these compounds is sufficient for use as an SOFC cathode. The nonzero charge transfer gap for LSM indicates this material may have some ionic conductivity issues, and it is experimentally known that the bulk oxygen conductivity of LSM is low.

**Table S2**. Tabulated experimental and predicted values for a set of well-studied SOFC cathode materials. The listed *k** and ASR data are for approximately T = 1000 K and $p(O_2)$=0.2 atm.

| Material | Experiment log *k** (cm/s) | Predicted log k* (cm/s) | Experiment log ASR (Ω-cm$^2$) | Predicted log ASR (Ω-cm$^2$) | Calculated stability (meV/atom above hull) | Experimental bulk stability sufficient for SOFC use?* | Calculated bandgap (eV) | Calculated charge transfer gap (eV) | Experimental electrical conductivity sufficient for SOFC use? |
|---|---|---|---|---|---|---|---|---|---|
| $La_{0.75}Sr_{0.25}MnO_3$ (LSM) | -8.75 | -8.32 | 2.05 | 2.03 | 32 | Yes | 0 | 0.18 | Yes |
| $La_{0.75}Sr_{0.25}CoO_3$ (LSC) | -6.70 | -7.23 | -0.08 | 0.42 | 45 | Yes | 0 | 0 | Yes |
| $La_{0.625}Sr_{0.375}Co_{0.25}Fe_{0.75}O_3$ (LSCF) | -6.01 | -5.83 | -0.59 | -0.12 | 47 | Yes | 0 | 0 | Yes |
| $Ba_{0.5}Sr_{0.5}Co_{0.75}Fe_{0.25}O_3$ (BSCF) | -3.71 | -3.37 | -1.69 | -1.93 | 124 | No | 0 | 0 | Yes |

*The activity of materials such as LSM and LSC is known to degrade over time due to cation segregation to the surface. Our stability screening is focused solely on bulk thermodynamic stability, so we ignore surface processes.

## 10. Promising materials that do not contain alkaline earth elements

**Table S3** is a summary of the promising materials in this work that pass all screening criteria as well as do not contain any alkaline earth elements. These materials may be potentially promising SOFC cathodes for designs seeking to eliminate alkaline earth elements as constituents.

**Table S3.** Summary of material compositions that pass all screening criteria and do not contain alkaline earth elements.

| Material Composition | Calculated O *p*-band center (eV) | Predicted log k* (cm/s) | Calculated energy above convex hull (meV/atom) |
|---|---|---|---|
| $YFe_{0.875}Ni_{0.125}O_3$ | -1.870 | -4.5023 | 37.5 |
| $Pr_{0.5}Nd_{0.5}CoO_3$ | -2.027 | -5.0276 | 25.3 |
| $Pr_{0.5}Sm_{0.5}CoO_3$ | -2.042 | -5.0778 | 35.1 |
| $Nd_{0.5}Y_{0.5}CoO_3$ | -2.046 | -5.0912 | 45.4 |
| $PrFe_{0.875}Co_{0.125}O_3$ | -2.140 | -5.4057 | 45.5 |
| $Pr_{0.75}Nd_{0.25}CoO_3$ | -2.164 | -5.4860 | 25.0 |
| $PrFe_{0.125}Co_{0.875}O_3$ | -2.210 | -5.6399 | 42.2 |
| $PrCoO_3$ | -2.218 | -5.6667 | 22.4 |
| $YFe_{0.875}Co_{0.125}O_3$ | -2.238 | -5.7336 | 27.8 |

## 11. Competing stability of hexagonal and Ruddlesden-Popper materials

For the select set of compounds we predicted to have $k^*$ values on par with or higher than BSCF ($k^* \approx -3.8$ cm/s) in **Table 1** of the main text, we also simulated the hexagonal and Ruddlesden-Popper (n=1) phases, as these phases are often competitive in stability with the perovskites. We considered only this limited set of nine exceptionally promising compounds to keep the calculations tractable. Other compounds have not been screened for stability against the hexagonal and Ruddlesden-Popper (n=1) phases as they are not standardly in the Materials Project database we used for determining stability. **Figure S5** provides a summary of the calculated stability under ORR operating conditions for the perovskite, Ruddlesden-Popper and hexagonal phases of these top-performing materials. Overall, we found for each case that the perovskite phase was still stable (i.e., not pushed above the 47 meV/atom stability criterion) by the presence of the Ruddlesden-Popper phase. On the other hand, some hexagonal materials were nearly degenerate in energy with their perovskite variants. Due to the near-degeneracy of perovskite and hexagonal structures for some compounds, synthesis of these compounds may result in either perovskite, hexagonal, or a mixture these phases. However, the perovskite phase might be isolated by careful control of synthesis conditions and/or with post-synthesis processing steps such as high temperature heat treatment. For example, such changes for processing BSCF have been shown to modify the amount of secondary phases present by controlling the grain size.[46, 47] Overall, the exact experimental methods needed to produce phase-pure perovskite materials will be a function of the material composition considered.

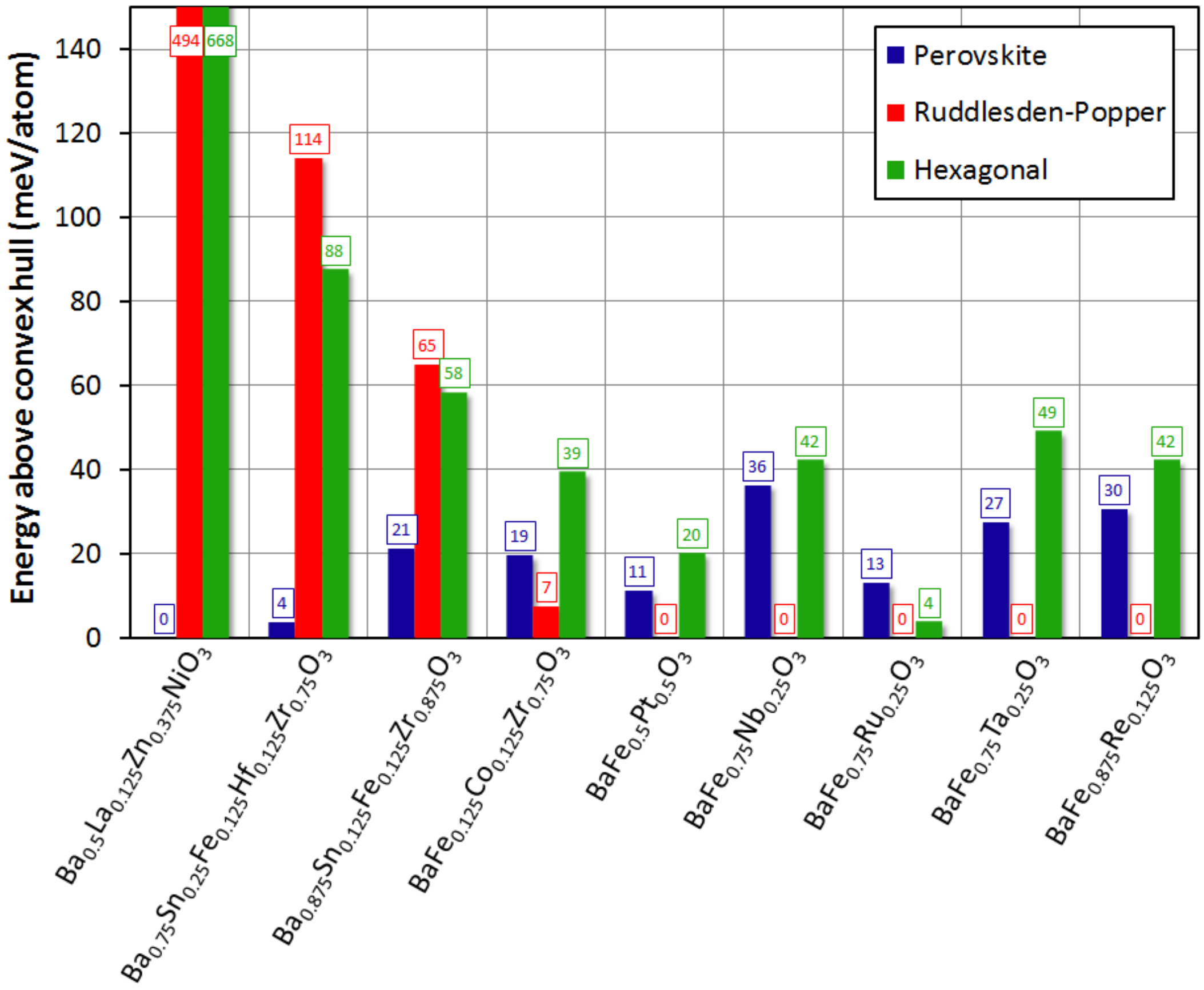

**Figure S5.** Comparison of calculated stability between perovskite, Ruddlesden-Popper and hexagonal phases. The materials listed here are a selected subset of the highest $k^*$ materials predicted from our screening process as summarized in **Table 1** of the main text and this **SI**.

## 12. Stability of promising materials with solid electrolytes

For a subset of ten of the top compounds (plus LSM as a material to compare with literature results) listed in **Table 1** of the main text we have assessed their stability in the presence of common solid electrolyte materials. This set of ten materials was chosen for this analysis because they represent some of the most promising materials predicted in this study, and also contain a spectrum of different doping elements and doping site fractions. For this analysis, we performed the same phase stability calculations as in **Section 2.3** of the main text, but this time assumed a 50/50 compositional mixture of cathode material and electrolyte material, which is an approximate representation of the material composition at the interface of the cathode and electrolyte. The common electrolyte materials we have considered here are yttria-stabilized zirconia (YSZ), gadolinium-doped ceria (GDC) and $Bi_2O_3$. The specific electrolyte compositions used in this analysis are: YSZ: $0.1Y_2O_3$-$0.9ZrO_2$, GDC: $Ce_{0.8}Gd_{0.2}O_{1.9}$, $Bi_2O_3$: pure $Bi_2O_3$.[48] It is well-known that the cathode material LSM is unstable with YSZ, but stable with GDC.[49] We have used the stability of the LSM + YSZ and LSM + GDC systems as a benchmark to gauge the expected stability of the new compounds with these electrolytes. A summary of the stability of each cathode + electrolyte system is given in **Table S4**. The stability of each system was assessed qualitatively by whether or not the decomposition products of the cathode + electrolyte mixture resulted in a reaction between the cathode and electrolyte. For example, in the case of cathode + YSZ, if there were decomposition products containing Y or Zr (besides the $Y_2O_3$ or $ZrO_2$ which comprised the electrolyte), the system was deemed unlikely to be stable because the cathode and electrolyte reacted to form other phases. Due to the frequent absence of total energies for solid solutions in the Materials Project database, solid solutions are usually not explicitly calculated but instead treated as mixtures of their line-compound endmembers, with a mixing energy of zero. Formally, the way we model the cathode + electrolyte stability is to compare the energy of the initial mixture to possible decomposition products and find the change in energy to the most stable decomposition, which could be just remaining in the initial stable state if the system is stable. For example, to determine the stability of 50% LSM and 50% YSZ we considered a mixture at the 50/50 point in the composition space, and assigned an energy to it that is the 50/50 weighted average of bulk LSM and YSZ (represented as bulk $Y_2O_3$ + $ZrO_2$). Encouragingly, our analysis of LSM + YSZ and LSM + GDC reproduces the well-known fact that LSM is not stable with YSZ, and is stable with GDC. Thus, while the stability analysis performed here is highly qualitative and approximate, this analysis should still provide useful guidance to choose the appropriate electrolyte(s) to pair with our newly predicted cathode materials.

By examining the stability of our new cathode materials with each electrolyte, we found that $Ba_{0.5}La_{0.125}Zn_{0.375}NiO_3$ is the only cathode material expected to be compatible with YSZ, GDC, and $Bi_2O_3$. However, we note that no kinetic considerations are included here, which might enable even unstable phases to coexist for long periods of time. In addition, minor compositional engineering can be used to reduce reactivity. For example, LSM+YSZ is predicted to be unstable, and $La_2Zr_2O_7$ has been observed at electrode/electrolyte interfaces of these materials[50, 51] However, the formation of this phase is also known to be suppressed by using cation-deficient LSM, where the La deficiency presumably reduces the La activity in the LSM and destabilizes the formation of the La containing $La_2Zr_2O_7$.[52-54] Finally, we note that instability can be counteracted by using barrier layers.[55-58] As all predicted compounds (except $Ba_{0.5}La_{0.125}Zn_{0.375}NiO_3$) at least partially react with these common electrolytes, specific design strategies using thin buffer layers, alternative electrolyte compositions, or cation-deficient cathodes may be necessary to sufficiently stabilize these promising cathode materials in a full SOFC device.

**Table S4.** Summary of stability analysis of LSM and top predicted cathode materials with commonly used solid oxide electrolytes yttria-stabilized zirconia (YSZ), gadolinium-doped ceria (GDC) and $Bi_2O_3$. The designation of whether the cathode + electrolyte system is expected to react is explained in the main text.

| **Electrolyte = YSZ** | | | |
|---|---|---|---|
| **Cathode material** | **Energy above convex hull (meV/atom)** | **Cathode + Electrolyte (50/50 mixture) decomposition products** | **Cathode and electrolyte expected to react?** |
| $La_{0.75}Sr_{0.25}MnO_3$ (LSM) | 35.3 | $LaMn_2O_5$, $YMnO_3$, $SrZrO_3$, $Sr_2Zr_7O_{16}$, $La_2Zr_2O_7$, $Mn_3O_4$ | Yes |
| $BaFe_{0.875}Al_{0.125}O_3$ | 102.9 | $O_2$, $Fe_2O_3$, $Ba(FeO_2)_2$, $BaZrO_3$, $YFeO_3$, $BaAl_2O_4$ | Yes |
| $Ba_{0.75}Sr_{0.25}Fe_{0.875}Ga_{0.125}O_3$ | 94.6 | $Sr_3Ga_4O_9$, $BaZrO_3$, $SrZrO_3$, $YFeO_3$, $Fe_2O_3$, $O_2$, $Sr_2Fe_2O_5$ | Yes |
| $BaFe_{0.75}Ta_{0.25}O_3$ | 77.3 | $Fe_2O_3$, $Ba_5Ta_4O_{15}$, $O_2$, $BaZrO_3$, $YTaO_4$ | Yes |
| $SrCo_{0.5}Sc_{0.5}O_3$ | 90.2 | $SrZrO_3$, $O_2$, $Co_3O_4$, $Sr_2Co_2O_5$, $Y_2O_3$, $Sc_2O_3$ | Yes |
| $BaFe_{0.125}Co_{0.125}Zr_{0.75}O_3$ | 30.5 | $O_2$, $BaZrO_3$, $Y_4Zr_3O_{12}$, $Co_3O_4$, ZrO2, $Fe_2O_3$ | Yes |
| $BaFe_{0.875}Re_{0.125}O_3$ | 85.6 | $O_2$, $Ba_5Re_3O_{16}$, $Y_4Zr_3O_{12}$, $YFeO_3$, $BaZrO_3$, $Fe_2O_3$ | Yes |
| $Ba_{0.5}La_{0.125}Zn_{0.375}NiO_3$ | 0 | same as reactants | No |
| $Ba_{0.875}Sn_{0.125}Fe_{0.125}Zr_{0.875}O_3$ | 18.8 | $FeO$, $BaZrO_3$, $ZrO_2$, $Y_2Sn_2O_7$, $Y_4Zr_3O_{12}$ | Yes |
| $BaFe_{0.75}Ru_{0.25}O_3$ | 66.1 | $YFeO_3$, $O_2$, $BaZrO_3$, $Ba_2YRuO_6$, $Ba_2Ru_7O_{18}$, $Fe_2O_3$ | Yes |
| $SrFe_{0.875}Al_{0.125}O_3$ | 64.4 | $Sr_2Fe_2O_5$, $SrAl_2O_4$, $YFeO_3$, $O_2$, $Fe_2O_3$, $SrZrO_3$ | Yes |

| Electrolyte = GDC | | | |
|---|---|---|---|
| **Cathode material** | **Energy above convex hull (meV/atom)** | **Cathode + Electrolyte (50/50 mixture) decomposition products** | **Cathode and electrolyte expected to react?** |
| $La_{0.75}Sr_{0.25}MnO_3$ (LSM) | 20.3 | $LaMnO_3$, $CeO_2$, $Gd_2O_3$, $Sr_3La_5Mn_8O_{24}$ | No |
| $BaFe_{0.875}Al_{0.125}O_3$ | 43.7 | $O_2$, $Ba(FeO_2)_2$, $BaCeO_3$, $GdFeO_3$, $CeO_2$, $BaAl_2O_4$ | Yes |
| $Ba_{0.75}Sr_{0.25}Fe_{0.875}Ga_{0.125}O_3$ | 39.6 | $BaCeO_3$, $O_2$, $CeO_2$, $Sr_2Fe_2O_5$, $Ba(FeO_2)_2$, $GdFeO_3$, $Ba_3Ga_4O_9$ | Yes |
| $BaFe_{0.75}Ta_{0.25}O_3$ | 29.3 | $GdFeO_3$, $CeO_2$, $O_2$, $Ba(FeO_2)_2$, $BaCeO_3$, $Ba_5Ta_4O_{15}$ | Yes |
| $SrCo_{0.5}Sc_{0.5}O_3$ | 33.1 | $Sc_2O_3$, $Sr_2CeO_4$, $CeO_2$, $Gd_2O_3$, $Sr_2Co_2O_5$, $O_2$ | Yes |
| $BaFe_{0.125}Co_{0.125}Zr_{0.75}O_3$ | 13.5 | $O_2$, $GdFeO_3$, $BaZrO_3$, $Ba_2CoO_4$, $CeO_2$, $Gd_2O_3$ | Yes |
| $BaFe_{0.875}Re_{0.125}O_3$ | 32.7 | $CeO_2$, $BaCeO_3$, $Ba(FeO_2)_2$, $O_2$, $Ba_5Re_3O_{16}$, $GdFeO_3$ | Yes |
| $Ba_{0.5}La_{0.125}Zn_{0.375}NiO_3$ | 0 | same as reactants | No |
| $Ba_{0.875}Sn_{0.125}Fe_{0.125}Zr_{0.875}O_3$ | 19.9 | $FeO$, $Gd_2Sn_2O_7$, $Gd_2O_3$, $BaZrO_3$, $CeO_2$ | Yes |
| $BaFe_{0.75}Ru_{0.25}O_3$ | 11.6 | $O_2$, $Ba_3CeRu_2O_9$, $CeO_2$, $GdFeO_3$, $BaCeO_3$, $Ba(FeO_2)_2$ | Yes |
| $SrFe_{0.875}Al_{0.125}O_3$ | 21.1 | $Gd_2O_3$, $CeO_2$, $SrAl_2O_4$, $O_2$, $Sr_2CeO_4$, $Sr_2Fe_2O_5$ | Yes |
| **Electrolyte = $Bi_2O_3$** | | | |
| **Cathode material** | **Energy above convex hull (meV/atom)** | **Cathode + Electrolyte (50/50 mixture) decomposition products** | **Cathode and electrolyte expected to react?** |
| $La_{0.75}Sr_{0.25}MnO_3$ (LSM) | 16.1 | $Bi_2O_3$, $LaMnO_3$, $Sr_3La_5Mn_8O_{24}$ | No |
| $BaFe_{0.875}Al_{0.125}O_3$ | 54.9 | $BaBiO_3$, $O_2$, $BaAl_2O_4$, $Fe_2O_3$, $Bi_2O_3$ | Yes |
| $Ba_{0.75}Sr_{0.25}Fe_{0.875}Ga_{0.125}O_3$ | 48.4 | $BaBiO_3$, $O_2$, $Sr(BiO_2)_2$, $Fe_2O_3$, $Sr_3Ga_4O_9$, $Bi_2O_3$ | Yes |
| $BaFe_{0.75}Ta_{0.25}O_3$ | 36.2 | $Fe_2O_3$, $Bi_2O_3$, $Ba_2TaBiO_6$, $BaBiO_3$ | Yes |
| $SrCo_{0.5}Sc_{0.5}O_3$ | 52.7 | $Co_3O_4$, $O_2$, $Sr(BiO_2)_2$, $Sc_2O_3$ | Yes |
| $BaFe_{0.125}Co_{0.125}Zr_{0.75}O_3$ | 14.7 | $Bi_2O_3$, $BaZrO_3$, $Co_3O_4$, $BaBiO_3$, $Fe_2O_3$, $O_2$ | Yes |
| $BaFe_{0.875}Re_{0.125}O_3$ | 36.1 | $BaBiO_3$, $Ba_5Re_3O_{16}$, $Bi_2O_3$, $Fe_2O_3$, $Ba(FeO_2)_2$ | Yes |
| $Ba_{0.5}La_{0.125}Zn_{0.375}NiO_3$ | 0 | same as reactants | No |
| $Ba_{0.875}Sn_{0.125}Fe_{0.125}Zr_{0.875}O_3$ | 12.9 | $Fe_3O_4$, Bi, $Sn_2Bi_2O_7$, $Bi_2O_3$, $BaZrO_3$ | Yes |
| $BaFe_{0.75}Ru_{0.25}O_3$ | 22.4 | $BaRuO_3$, $Fe_2O_3$, $Bi_2O_3$, $BaBiO_3$ | Yes |
| $SrFe_{0.875}Al_{0.125}O_3$ | 28.9 | $Sr(BiO_2)_2$, $O_2$, $SrAl_2O_4$, $Fe_2O_3$, $Bi_2O_3$ | Yes |